\documentclass[aps,reprint,
superscriptaddress, amsmath,amssymb, prd, longbibliography, twocolumn, floatfix
]{revtex4-2}
\usepackage{graphicx}% Include figure files
\usepackage{dcolumn}% Align table columns on decimal point
\usepackage{bm}% bold math
\usepackage{float}
 \usepackage{booktabs}
\usepackage[caption=false]{subfig}
\usepackage{amsmath}
\usepackage{comment}
\usepackage{calligra}
\DeclareMathAlphabet{\mathcalligra}{T1}{calligra}{m}{n}
\DeclareFontShape{T1}{calligra}{m}{n}{<->s*[2.2]callig15}{}

\usepackage{amssymb}
\usepackage{tabularx}
\usepackage{booktabs}
\usepackage{multirow}  % needed for some tables
\usepackage{xcolor}
\usepackage{natbib}
\usepackage{placeins}
\usepackage{slashed}

\usepackage{url}
\usepackage{fontawesome5}
\usepackage{scalerel}
\usepackage{tikz}
\usetikzlibrary{svg.path} 
\definecolor{orcidlogocol}{HTML}{A6CE39}
\tikzset{
  orcidlogo/.pic={
    \fill[orcidlogocol] svg{M256,128c0,70.7-57.3,128-128,128C57.3,256,0,198.7,0,128C0,57.3,57.3,0,128,0C198.7,0,256,57.3,256,128z};
    \fill[white] svg{M86.3,186.2H70.9V79.1h15.4v48.4V186.2z}
                 svg{M108.9,79.1h41.6c39.6,0,57,28.3,57,53.6c0,27.5-21.5,53.6-56.8,53.6h-41.8V79.1z M124.3,172.4h24.5c34.9,0,42.9-26.5,42.9-39.7c0-21.5-13.7-39.7-43.7-39.7h-23.7V172.4z}
                 svg{M88.7,56.8c0,5.5-4.5,10.1-10.1,10.1c-5.6,0-10.1-4.6-10.1-10.1c0-5.6,4.5-10.1,10.1-10.1C84.2,46.7,88.7,51.3,88.7,56.8z};
  }
}

\newcommand\orcidicon[1]{\href{https://orcid.org/#1}{\mbox{\scalerel*{
\begin{tikzpicture}[yscale=-1,transform shape]
\pic{orcidlogo};
\end{tikzpicture}
}{|}}}}

\definecolor{mycolor}{RGB}{0,0,204}
\definecolor{pink}{RGB}{255,0,127}

\usepackage[unicode]{hyperref}
\hypersetup{
  colorlinks   = true,    % Colours links instead of ugly boxes
  urlcolor     = blue,    % Colour for external hyperlinks
  linkcolor    =pink,    % Colour of internal links
  citecolor    = pink     % Colour of citations
}

\usepackage{tabularx}
\usepackage{graphicx}% Include figure files
\usepackage{dcolumn}% Align table columns on decimal point
\usepackage{bm}% bold math

\begin{document}

\preprint{APS/123-QED}
\title{Non-Minimal Dilaton Inflation From The Effective Gluodynamics } 

 \author{Pirzada %\orcidicon{0009-0002-2274-9218}
 }%
 %\email{pirzada@itp.ac.cn}%
\affiliation{CAS Key Laboratory of Theoretical Physics, Institute of Theoretical Physics, Chinese Academy of Sciences, Beijing 100190, China}
\affiliation{School of Physical Sciences, University of Chinese Academy of Sciences, No. 19A Yuquan Road, Beijing 100049, China}
\author{Imtiaz Khan}
\email{ikhanphys1993@gmail.com}
\affiliation{Department of Physics, Zhejiang Normal University, Jinhua, Zhejiang 321004, China}
\affiliation{Research Center of Astrophysics and Cosmology, Khazar University, Baku, AZ1096, 41 Mehseti Street, Azerbaijan}
%\affiliation{Zhejiang Institute of Photoelectronics, Jinhua, Zhejiang 321004, China}
 \author{Mussawir Khan}%
\affiliation{State Key Laboratory of Particle Astrophysics, Institute of High Energy Physics, Chinese Academy of Sciences, Beijing 100049, China}
\affiliation{University of Chinese Academy of Sciences, Beijing 100049, China}
  \author{Tianjun Li }%
 %\email{tli@itp.ac.cn}%
 %\affiliation{CAS Key Laboratory of Theoretical Physics, Institute of Theoretical Physics, Chinese Academy of Sciences, Beijing 100190, China }
 \affiliation{School of Physics, Henan Normal University, Xinxiang 453007, P. R. China}%
\author{Ali Muhammad}
%\email{alimuhammad@phys.qau.edu.pk}
\affiliation{CAS Key Laboratory of Theoretical Physics, Institute of Theoretical Physics, Chinese Academy of Sciences, Beijing 100190, China}
\affiliation{School of Physical Sciences, University of Chinese Academy of Sciences, No. 19A Yuquan Road, Beijing 100049, China}

%\date{\today}
\begin{abstract}

{We develop a nonminimal dilaton-inflation model in which the inflaton is the lightest scalar excitation of a hidden confining gauge theory. The Migdal--Shifman anomaly-matching action fixes a logarithmic contribution to the scalar potential, $V_{\rm MS}=A\varphi^4[\ln(\varphi/\mu)-1/4]$, with $(A,\mu)$ mapped to the scalar mass and vacuum condensate. Embedding this sector in the leading curved-space EFT introduces the independent Wilson coefficients $\lambda$ and $\xi$, and the resulting Einstein-frame dynamics yields a plateau with a calculable anomaly-induced deformation. We analyze the pure MS limit as a baseline, derive the $\mu$--$\lambda$ reparametrization, state finite-window RG-control conditions, and impose both the nonminimal-gravity cutoff and the intrinsic confining-sector gap. Exact slow-roll scans show that the viable regime combines the usual large-$\xi$ attractor behavior with a logarithmic imprint tied directly to nonperturbative trace-anomaly matching.}
\end{abstract}

\maketitle

% ==========================================================
\section{Introduction}
% ==========================================================

The inflationary paradigm \cite{Starobinsky:1980,Guth:1981,Linde:1983} provides a unified dynamical resolution of the horizon, flatness, and relic problems of hot Big Bang cosmology, while offering a quantum origin for the primordial perturbations that seed structure. Precision CMB temperature and polarization data now constrain inflationary dynamics quantitatively: the Planck 2018 legacy results and recent ACT/DESI data tightly determine the scalar tilt and amplitude and significantly restrict viable single-field slow-roll realizations \cite{Planck:2018,ACTDR6LCDM:2025,ACTDR6Ext:2025,Abdul_Karim_2025,Ijaz:2024zma,Khan:2023snv,Ijaz:2023cvc}. In parallel, B-mode searches by BICEP/Keck and other ground-based experiments yield stringent bounds on the tensor-to-scalar ratio \cite{BK:2024,2022,Sayre_2020}, sharpening the empirical preference for concave, plateau-like Einstein-frame potentials. These constraints motivate constructions in which the required flatness is supported by robust structural principles, notably symmetries, anomalies, and controlled infrared effective descriptions of strong dynamics.

A natural organizing principle is (approximate) scale invariance and its breaking \cite{Callan:1970}. While classical scale invariance is generic in renormalizable theories without explicit mass parameters, it is typically violated by renormalization-group running\cite{Komargodski_2011,Elvang_2013,Einhorn_2015}. In asymptotically free non-Abelian gauge theories, dimensional transmutation generates a physical scale even when the ultraviolet Lagrangian contains no dimensionful couplings. This quantum breaking is encoded in the trace anomaly through the non-vanishing trace of the renormalized energy--momentum tensor, $\theta \equiv \theta^\mu{}_\mu$, and its vacuum expectation value. The anomaly further implies an infinite tower of low-energy theorems \cite{Crewther:1972kn,Corian__2013,Yonekura_2010,thuorst2024lowenergytheoremslinearitybreaking,schwimmer2023commentstraceanomalymatching} (integrated Ward identities) for correlation functions of $\theta$ at vanishing external momenta \cite{Collins:1976yq}. Any EFT for the lightest scalar excitation in this channel must reproduce these constraints, making anomaly matching a sharp infrared target for model building.

A particularly economical realization of this logic is the Migdal--Shifman (MS) effective theory of gluodynamics \cite{Migdal:1982}. In this construction a single dimensionless scalar(Dilaton) field $X$, representing the lightest $0^{++}$ gluonic mode, is introduced such that the improved trace operator becomes an exponential functional on-shell, $\theta \propto e^X$. This exponential form is precisely what is required to saturate the full tower of trace-anomaly Ward identities already at tree level. After a canonical reparametrization in $D=4$, the resulting non-polynomial MS dynamics yield a Coleman--Weinberg--type potential $V(\varphi)\propto \varphi^4\!\left(\ln(\varphi/\mu)-1/4\right)$, closely paralleling radiatively generated symmetry-breaking structures \cite{ColemanWeinberg:1973}. In the MS case, the logarithmic dependence has a distinct origin: it is dictated by infrared anomaly matching instead of a perturbative loop expansion.

The cosmological relevance of anomaly-induced scalar dynamics has been explored in composite/glueball inflation scenarios, where the inflaton is identified with the lightest scalar of a confining gauge sector \cite{Bezrukov:2012prd}. A key mechanism is the non-minimal coupling $\xi \varphi^2 R$, which can flatten the Einstein-frame potential: for sufficiently large $\xi$, Jordan-frame potentials that grow quartically at large field approach an asymptotically flat plateau, enabling slow-roll evolution consistent with CMB bounds. This is the basis of Higgs inflation and related models and is closely connected to strong-coupling attractor behavior \cite{Bezrukov:2008,Kallosh:2014,Ijaz:2024zma}. In these regimes the leading predictions approach $n_s\simeq 1-2/N$ and $r\simeq 12/N^2$, up to controlled corrections that are largely insensitive to microscopic details.

{The inflationary model constructed below has two logically distinct ingredients. The first is the Migdal--Shifman anomaly sector: the logarithmic contribution and its normalization map to $(m,|e_{\rm vac}|)$ are fixed by trace-anomaly matching. The second is the gravitational EFT completion: the quartic coefficient $\lambda$ and nonminimal coupling $\xi$ are independent renormalized Wilson coefficients. The resulting claim is therefore precise and testable: once the leading plateau dynamics is generated by the nonminimal coupling, the dominant logarithmic departure from the attractor has a nonperturbative origin and a definite matching to the confining sector. This separates the present construction from phenomenological running-inflation potentials in which the coefficient of $\varphi^4\ln\varphi$ is treated as a free beta-function parameter \cite{lee1997runninginflation,De_Simone_2009,M_ri_n_2020}. It also separates it from generic composite-inflation parametrizations by retaining the exact Migdal--Shifman mapping between $(A,\mu)$ and $(m,|e_{\rm vac}|)$ \cite{Migdal:1982,Pirzada:2026sle,Bezrukov:2012prd}.}

{This distinction matters phenomenologically. A shift of $\mu$ in a Coleman--Weinberg-like potential can be traded against $\lambda$ unless $\mu$ is fixed by a microscopic matching prescription. Here $\mu=4\sqrt{|e_{\rm vac}|}/m$ is part of the MS matching, while $\lambda$ and $\xi$ are EFT parameters to be specified at the inflationary matching scale. We therefore make the parameter status explicit, present the pure MS limit before adding the gravitational completion, quantify the domain in which RG running of $\lambda$ and $\xi$ may be consistently neglected over the CMB window, and impose both the background-dependent nonminimal-coupling cutoff and the intrinsic confining-sector cutoff associated with heavier glueball states.}

{The paper is organized as follows. In Sec.~\ref{sec:MS} we review the MS gluodynamics EFT and derive the canonical logarithmic potential from the original nonpolynomial Lagrangian. In Sec.~\ref{sec:gravity} we construct the gravitational completion, identify the independent Wilson coefficients, and display the $\mu$--$\lambda$ reparametrization. In Sec.~\ref{sec:inflation} we derive the exact slow-roll machinery and give the pure-MS baseline before the large-$\xi$ attractor analysis. In Sec.~\ref{numerics} we present reproducible numerical scans, including the pure-MS diagnostic plots and the nonminimal plateau results. In Sec.~\ref{sec:discussion} we analyze EFT control, including the intrinsic confining-sector gap. We conclude in Sec.~\ref{sec:conclusion}.}

% ==========================================================
\section{Migdal--Shifman gluodynamics and anomaly-matching scalar EFT}
\label{sec:MS}
% ==========================================================
The starting point of Migdal and Shifman (MS) is the observation that in an asymptotically free theory with dimensional transmutation, the trace of the (regularized) energy--momentum tensor,
\begin{equation}
\theta \equiv \theta^\mu{}_\mu,
\end{equation}
is a distinguished operator that controls how the vacuum energy and correlators respond to rescalings. In pure Yang--Mills (gluodynamics), \(\theta\) is fixed by the beta function,
\begin{equation}
\theta = \frac{\beta(g)}{2g}\, G^a_{\mu\nu}G^{a\,\mu\nu},
\end{equation}
up to scheme-dependent contact terms. Migdal and Shifman emphasize that the trace anomaly implies an \emph{infinite tower} of low-energy theorems (integrated Ward identities) for connected correlators with insertions of \(\theta\). A representative form of this tower in \(D\) dimensions is (\textit{We follow the MS normalization in which the right-hand side is proportional to \(( -D )^n\) times the vacuum condensate; see} \cite{Migdal:1982}.)

\begin{equation}
\begin{split}
i^n \int d^D x_1\cdots d^D x_n\,
\big\langle 0\big|T\{\theta(x_1)\cdots \theta(x_n)\theta(0)\}\big|0\big\rangle_{\rm conn} \\
= \langle \theta\rangle_{\rm vac} \,(-D)^n,
\end{split}
\label{eq:MSWard}
\end{equation}
valid at zero external momenta after appropriate subtraction of contact terms.
Equation \eqref{eq:MSWard} is extremely constraining: it fixes the entire hierarchy of zero-momentum amplitudes of the scalar channel associated with \(\theta\). MS propose to build a \emph{single-field} effective theory for the lightest \(0^{++}\) excitation such that \eqref{eq:MSWard} is saturated already at tree level.

The key idea is that a one-scalar EFT can reproduce \eqref{eq:MSWard} if the trace operator \(\theta\) becomes an \emph{exponential} of the effective field. Exponentials generate factorial structures under repeated differentiation with respect to sources, matching the repeated insertions in the Ward identities. MS therefore seek an EFT in which the improved stress tensor has a trace \(\theta \propto e^{X}\) upon using the scalar equation of motion.

\textit{\textbf{{ Migdal--Shifman effective Lagrangian}}}

Migdal and Shifman introduce a dimensionless scalar(Dilaton) field \(X\) representing the lightest scalar gluonic mode and propose the effective Lagrangian in \(D\) dimensions (their Eqs.~(8)--(9)):
\begin{equation}
\mathcal{L}_{\rm MS}
=
\frac{|e_{\rm vac}|}{m^{2}}\,
\frac12(\partial_\mu X)^2\, e^{\frac{D-2}{D}X}
-
V_{\rm MS}(X),
\label{eq:LMS}
\end{equation}
with potential
\begin{equation}
V_{\rm MS}(X)=|e_{\rm vac}|\,(X-1)\,e^{X}.
\label{eq:VMSX}
\end{equation}
\begin{figure}[!ht]
\centering
\includegraphics[width=0.92\linewidth]{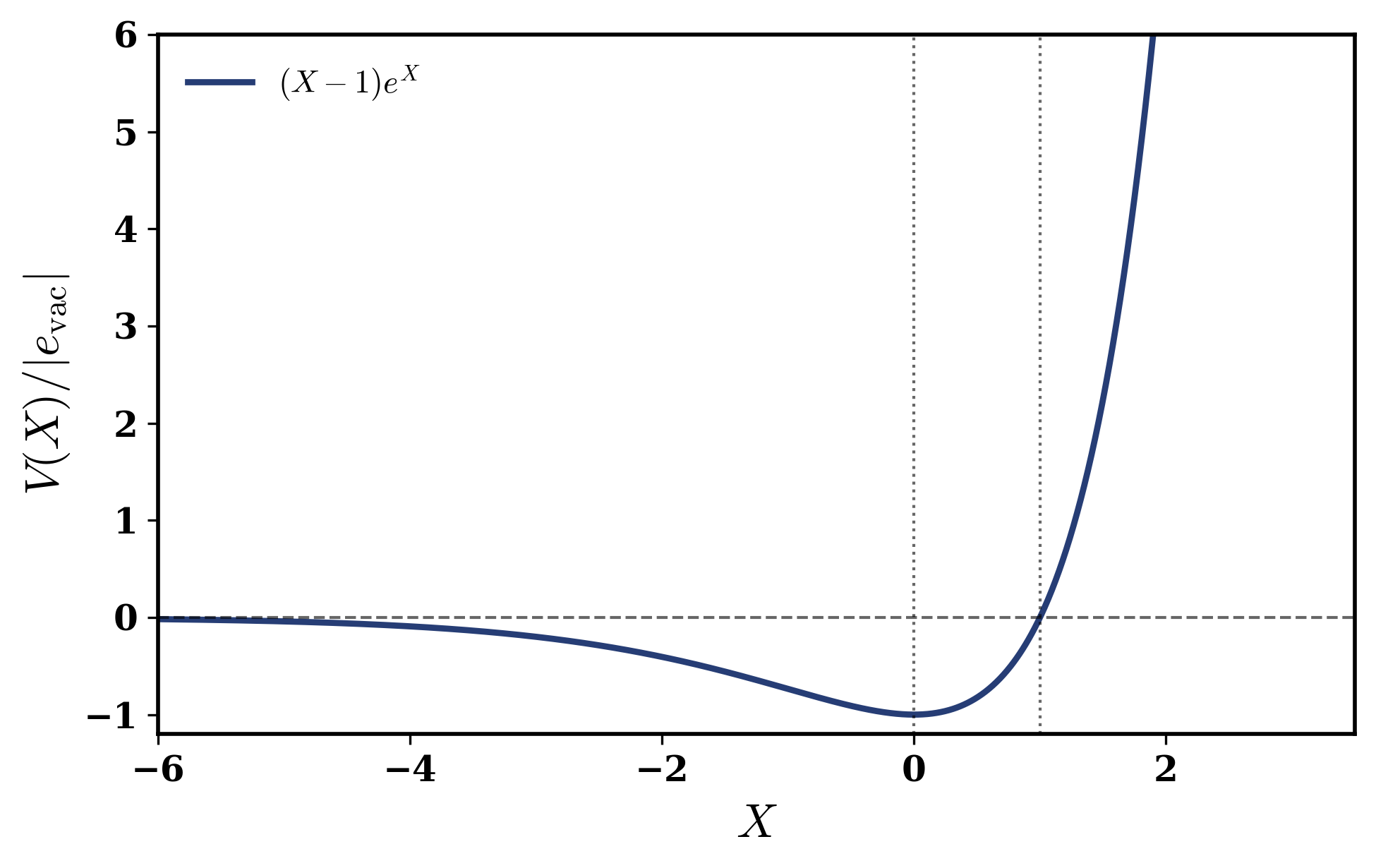}
\caption{MS potential in \(X\). \(V(X)/|e_{\rm vac}|=(X-1)e^X\) \cite{Migdal:1982}.}
\end{figure}
Here \(m\) is the mass scale of the scalar excitation, while \(e_{\rm vac}<0\) is the (negative) vacuum energy density of the confining theory, so \(|e_{\rm vac}|\equiv -e_{\rm vac}>0\).
The sign \(e_{\rm vac}<0\) is required for the existence of a stable one-meson realization of the Ward identities \cite{Migdal:1982,appelquist2022dilatoneffectivefieldtheory,Kamath_2006}.

The MS construction is not a random Ansatz: it is engineered so that the improved stress tensor has the correct anomaly structure. Indeed, the potential \eqref{eq:VMSX} is precisely chosen so that when the equation of motion is used, the trace operator becomes proportional to \(e^X\). One convenient way to see this is to note that scale transformations shift \(X\) by a constant and rescale the metric, and the improved trace picks up a contribution proportional to the variation of the action under this shift. MS show that, on-shell,
\begin{equation}
\theta \;\propto\; -D\,|e_{\rm vac}|\,e^{X},
\label{eq:thetaExp}
\end{equation}
so that repeated insertions of \(\theta\) are captured by repeated derivatives of an exponential, reproducing the hierarchy \eqref{eq:MSWard} at tree level. Equation \eqref{eq:thetaExp} is the precise sense in which the MS EFT packages the infinite tower of Ward identities into a one-field description.

\textit{\textbf{{Canonical variable and the \(\varphi^4\ln\varphi\) potential in \(D=4\)}}}

For inflationary applications we work in \(D=4\), where the kinetic prefactor in \eqref{eq:LMS} becomes \(e^{X/2}\). A simple field redefinition makes the kinetic term polynomial. Define
\begin{equation}
\begin{split}
 & \chi \equiv e^{X/4}  \\&
\qquad\Longleftrightarrow\qquad
X = 4\ln\chi,\quad e^{X/2}=\chi^{2},\quad e^X=\chi^4~.
\end{split}
\label{eq:chiDef}
\end{equation}
Then
\begin{equation}
\begin{split}
& \partial_\mu X = \frac{4}{\chi}\partial_\mu\chi
\\&
\quad\Rightarrow\quad
(\partial X)^2 e^{X/2} = \left(\frac{16}{\chi^2}(\partial\chi)^2\right)\chi^2
=16(\partial\chi)^2~.
\end{split}
\end{equation}
Substituting into \eqref{eq:LMS} at \(D=4\) yields

\begin{equation}
\begin{split}
\mathcal{L}_{\rm MS}^{(4)} &=
\frac{|e_{\rm vac}|}{m^{2}}\,\frac12\cdot 16\,(\partial\chi)^2
-
|e_{\rm vac}|(4\ln\chi-1)\chi^4 \\
&=
\frac{8|e_{\rm vac}|}{m^2}(\partial\chi)^2
-
|e_{\rm vac}|(4\ln\chi-1)\chi^4.
\end{split}
\label{eq:LMSchi}
\end{equation}
Now canonically normalize by defining a dimension-one field \(\varphi\) via
\begin{equation}
\varphi \equiv \frac{4\sqrt{|e_{\rm vac}|}}{m}\,\chi,
\qquad
\mu \equiv \frac{4\sqrt{|e_{\rm vac}|}}{m},
\qquad\Rightarrow\qquad
\chi=\frac{\varphi}{\mu}.
\label{eq:varphiDef}
\end{equation}
This rescaling is fixed uniquely by the requirement that the kinetic term takes the canonical form $\frac12(\partial\varphi)^2$ in four dimensions. Since $|e_{\rm vac}|$ has mass dimension $4$ and $m$ has mass dimension $1$, the prefactor $\sqrt{|e_{\rm vac}|}/m$ has mass dimension $1$, and therefore the canonically normalized field $\varphi$ correctly has mass dimension $1$ as required for a scalar in 4D.
With this choice,
\begin{equation}
\begin{split}
\frac{8|e_{\rm vac}|}{m^2}(\partial\chi)^2
=
\frac{8|e_{\rm vac}|}{m^2}\left(\frac{1}{\mu^2}\right)(\partial\varphi)^2 \\
=
\frac{8|e_{\rm vac}|}{m^2}\left(\frac{m^2}{16|e_{\rm vac}|}\right)(\partial\varphi)^2
=
\frac12(\partial\varphi)^2,
\end{split}
\end{equation}
so the kinetic term becomes canonical. The potential becomes
\begin{align}
V_{\rm MS}(\varphi)
&=
|e_{\rm vac}|\left(4\ln\frac{\varphi}{\mu}-1\right)\left(\frac{\varphi}{\mu}\right)^4
\nonumber\\
&=
\left(\frac{4|e_{\rm vac}|}{\mu^4}\right)\varphi^4\left(\ln\frac{\varphi}{\mu}-\frac14\right).
\label{eq:VMSvarphiDerive}
\end{align}
Defining
\begin{equation}
A \equiv \frac{4|e_{\rm vac}|}{\mu^4}
\qquad\Rightarrow\qquad
A=\frac{m^4}{64|e_{\rm vac}|},
\label{eq:AMap}
\end{equation}
we obtain the standard MS canonical form. 
\begin{equation}
V_{\rm MS}(\varphi)
=
A\,\varphi^4\left(\ln\frac{\varphi}{\mu}-\frac14\right),
\label{eq:VMS}
\end{equation}

with \(\mu\) and \(A\) related to \((m,|e_{\rm vac}|)\) exactly as in \eqref{eq:varphiDef} and \eqref{eq:AMap}. The potential has a minimum at \(\varphi=\mu\) and its value at the minimum reproduces the gluodynamics vacuum energy:
\begin{equation}
\left.\frac{dV_{\rm MS}}{d\varphi}\right|_{\mu}=0,
\qquad
V_{\rm MS}(\mu)=-\frac{A}{4}\mu^4=-|e_{\rm vac}|.
\end{equation}
{The scale $\mu = 4\sqrt{|e_{\rm vac}|}/m$ is a matching scale of the hidden trace-channel EFT rather than a QCD input. In an ordinary QCD-like theory with a single confinement scale $\Lambda_h$, one expects $m\sim\Lambda_h$ and $|e_{\rm vac}|^{1/4}\sim\Lambda_h$, giving $\mu/m=O(1)$. The inflationary benchmarks instead require a high-scale hidden sector in which this ratio is enhanced. Two standard mechanisms can realize such an enhancement. In a large-$N_h$ pure-gauge sector, planar counting gives $|e_{\rm vac}|\simeq c_e N_h^2\Lambda_h^4$ while glueball masses remain $m\simeq c_0\Lambda_h$ up to $1/N_h^2$ corrections \cite{tHooft:1974pnl,Witten:1979kh,Lucini:2004my,Lucini:2010nv}. The MS map then implies}
{
\begin{equation}
\frac{\mu}{m}\simeq \frac{4\sqrt{c_e}}{c_0^2}N_h,\qquad
A\simeq \frac{c_0^4}{64c_eN_h^2}.
\label{eq:largeNMSscaling}
\end{equation}}
{Thus large $\mu/m$ and a small anomaly coefficient are correlated rather than independent assumptions in the large-$N_h$ limit. A second possibility is a near-conformal hidden sector, where approximate scale symmetry can make the dilatonic scalar lighter than the heavier resonances; dilaton EFTs and lattice-motivated analyses of such walking theories provide precisely this separation mechanism \cite{Golterman:2016,Appelquist:2017,appelquist2022dilatoneffectivefieldtheory}. In the present work we do not choose a unique microscopic gauge group. Instead, Eqs.~\eqref{eq:AMap} and \eqref{eq:largeNMSscaling} identify the matching conditions that any hidden-sector realization must satisfy, while Sec.~\ref{sec:discussion} imposes the resulting gap condition quantitatively.}
We also emphasize that $V_{\rm MS}$ is not bounded above: as $\varphi\to\infty$, $V_{\rm MS}(\varphi)\sim A\varphi^4\ln(\varphi/\mu)\to +\infty$ for $A>0$.
This completes the mathematically explicit bridge from the original MS nonpolynomial Lagrangian \eqref{eq:LMS}--\eqref{eq:VMSX} to the canonical scalar EFT with a $\varphi^4\ln\varphi$ potential \eqref{eq:VMS}. Our work, which aligns with the composite inflation paradigm where the inflaton is identified as a glueball field \cite{Bezrukov:2012prd}, provides a concrete microphysical grounding through the derived relation \eqref{eq:AMap}, $A = m^4/(64|e_{\rm vac}|)$. In contrast to phenomenological models where such a logarithmic coefficient is a free parameter, here it is fundamentally determined by non-perturbative quantities of the confining gauge sector: the physical mass of the lightest scalar glueball ($m$) and the gluodynamics vacuum energy density ($|e_{\rm vac}|$). While the specific value of $m^4 / |e_{\rm vac}|$ for a hidden sector is a phenomenological input, it is in principle calculable via lattice methods (for known gauge groups like QCD, the scalar glueball mass lies in the range $1.0$--$4.0$ GeV \cite{Morningstar_1999}). This anchors the inflationary potential \eqref{eq:VMS} and its gravitational completion in the anomaly-matching structure of the Migdal--Shifman Lagrangian \cite{Migdal:1982}, ensuring they are not ad hoc constructs but are tightly constrained by the underlying strongly coupled dynamics.

{Two points are worth separating because they become central in the inflationary application. First, the MS relation fixes a trace-channel matching coefficient; it does not by itself fix the complete Wilsonian action obtained after coupling the hidden sector to gravity and integrating out heavier states. Second, the single-field EFT is controlled only below the first omitted resonance. In a pure Yang--Mills-like sector the masses of the lowest glueball excitations are set by the same confinement scale and are calculable nonperturbatively in lattice studies \cite{Morningstar_1999,Lucini:2004my,Lucini:2010nv}; in near-conformal or large-$N_h$ hidden sectors the spectrum and condensate hierarchy can differ, but the condition that the inflationary scales remain below the trace-channel gap remains mandatory. This is the logic used in Sec.~\ref{sec:discussion} when the MS parameters are converted into an intrinsic cutoff.}

\textit{\textbf{{Inflationary interpretation}}}

To use \(\varphi\) as the inflaton, we interpret it as the lightest scalar mode of a \emph{hidden} confining gauge sector whose confinement scale is far above QCD and high enough  to support inflation. The MS EFT is then understood as the leading term in a derivative expansion for this light scalar, constrained by anomaly matching. Coupling this EFT to gravity in a consistent way is the next step; crucially, because the MS construction relies on an improved stress tensor, the gravitational embedding naturally suggests a nonminimal coupling to curvature \cite{Callan:1970}.

% ==========================================================
\section{Nonminimal coupling and field dynamics}
\label{sec:gravity}
% ==========================================================

In curved space, the improvement of the scalar stress tensor corresponds to the presence of a \(\xi\varphi^2R\) operator \cite{Callan:1970}.  We stress that while improvement motivates the \emph{operator}, the value of $\xi$ is a renormalized coupling in the gravitational EFT and need not equal the special conformal value $\xi=1/6$ except in a particular free-field limit. {The operator is also required by renormalization of scalar field theory on curved backgrounds; even if absent at one scale, it is generally regenerated by loops once matter interactions are present \cite{BirrellDavies:1982,Buchbinder:1992,ParkerToms:2009}. Thus the use of a nonminimal coupling is not an extra symmetry assumption but part of the standard curved-space EFT basis.} Motivated by this (and by the ubiquity of nonminimal couplings under renormalization), we consider the Jordan-frame action
\begin{equation}
\begin{split}
S_J = \int d^4x\sqrt{-g}\left[
\frac12F(\varphi)R
-\frac12(\nabla\varphi)^2
- V_J(\varphi)
\right], \\
\qquad F(\varphi) = M_{\rm Pl}^2 + \xi\varphi^2.
\end{split}
\label{eq:JordanAction}
\end{equation}

{The Jordan-frame potential contains the anomaly-matching MS contribution \eqref{eq:VMS}. Once this scalar is embedded in a gravitational EFT, however, the leading marginal self-interaction consistent with the assumed $\varphi\to-\varphi$ symmetry is also allowed. We therefore write}
{
\begin{equation}
V_J(\varphi)=\frac{\lambda}{4}\varphi^4
+ A\,\varphi^4\left(\ln\frac{\varphi}{\mu}-\frac14\right)
+V_0,
\label{eq:VJfull}
\end{equation}}

{where $V_0$ fixes the late-time vacuum energy and does not affect the inflationary plateau. The status of the parameters is then unambiguous: $(A,\mu)$ belong to the anomaly-matched MS sector, while $(\lambda,\xi)$ are Wilson coefficients of the gravitational EFT. The logarithmic term is mandatory if the single scalar is to reproduce the trace-anomaly Ward identities; the quartic and nonminimal couplings are not predicted by pure gluodynamics and must be matched to the UV completion or treated phenomenologically at the inflationary scale. This is the sense in which the model is predictive: the plateau mechanism is the familiar nonminimal one, but the leading logarithmic deformation has a fixed anomaly origin rather than an arbitrary radiative ansatz.}

{This form also clarifies why the hierarchy $\lambda\gg A$ used in part of the scan is not a claim that loop effects associated with $\lambda$ are absent. In a Coleman--Weinberg model the logarithmic coefficient is itself an RG beta-function contribution \cite{ColemanWeinberg:1973,De_Simone_2009}; in the MS model $A$ is instead a low-energy theorem matching coefficient. A UV completion may generate additional running of $\lambda$ and $\xi$, but those effects are separate from the anomaly coefficient and are consistently subleading whenever the finite-window conditions in Eq.~\eqref{eq:RGcontrol} hold. If they do not hold, the fixed-coupling potential \eqref{eq:VJfull} should be replaced by its RG-improved version, and the present scan should be interpreted as the zeroth-order matching calculation.}

{It is useful to display the dependence on}
{
\begin{equation}
\alpha\equiv \frac{A}{\lambda},\qquad
V_J=\lambda\varphi^4\left[\frac14+\alpha\left(\ln\frac{\varphi}{\mu}-\frac14\right)\right]+V_0 .
\label{eq:alphadef}
\end{equation}}

{The small-deformation regime is the domain in which the anomaly-matched logarithm can be read as a calculable perturbation of the plateau over the CMB field interval. Larger values of $\alpha$ are included in the numerical scan and are constrained directly by $(n_s,r)$ and by EFT control.}

{\paragraph{Matching-scale reparametrization and running.}
The shift of the logarithmic scale can be absorbed into the quartic coefficient. Explicitly, for $\mu\to \mu e^{\delta}$,}
{
\begin{align}
\frac{\lambda}{4}\varphi^4
+A\varphi^4\left[\ln\frac{\varphi}{\mu e^{\delta}}-\frac14\right]
&=\frac{\lambda-4A\delta}{4}\varphi^4
+A\varphi^4\left[\ln\frac{\varphi}{\mu}-\frac14\right].
\label{eq:muLambdaShift}
\end{align}}

{Thus inflationary observables depend on the invariant combination sampled over the CMB window unless $\mu$ is fixed independently. Equivalently, one may work with the local quartic coefficient at the pivot,}
{
\begin{equation}
\lambda_{\rm loc}(\varphi_*)\equiv \lambda+4A\left(\ln\frac{\varphi_*}{\mu}-\frac14\right),
\label{eq:lambdaloc}
\end{equation}}
{together with the logarithmic slope coefficient $A$. The constant part of the logarithm can be absorbed into $\lambda_{\rm loc}$, but the derivative with respect to $\ln\varphi$ is proportional to $A$ and cannot be removed by the redefinition. In the present construction $\mu$ is not merely a subtraction convention: it is the MS matching scale $\mu=4\sqrt{|e_{\rm vac}|}/m$. Equation~\eqref{eq:muLambdaShift} is therefore used as a consistency check on parametrization, while Eq.~\eqref{eq:lambdaloc} identifies the physical pivot-scale quantities entering the slow-roll scan.}

{The same EFT viewpoint fixes how the running of $\lambda$ and $\xi$ is treated. The potential \eqref{eq:VJfull} is written at an inflationary matching scale $\mu_{\rm R}$ chosen near the CMB field value. Neglecting the running of $\lambda$ and $\xi$ over the observable window is self-consistent when}
{
\begin{equation}
|\beta_\lambda|\,|\Delta\ln\varphi| \ll |\lambda_{\rm eff}|,\qquad
|\beta_\xi|\,|\Delta\ln\varphi| \ll \xi,
\label{eq:RGcontrol}
\end{equation}}

{where $\lambda_{\rm eff}/4\equiv \lambda/4+A[\ln(\varphi_*/\mu)-1/4]$ and $\Delta\ln\varphi$ is the field interval corresponding to the CMB range $N\simeq50$--$60$. In the large-$\xi$ plateau regime $y_N\equiv \xi\varphi_N^2/M_{\rm Pl}^2\simeq4N/3$, hence}
{
\begin{equation}
\Delta\ln\varphi_{50-60}
\simeq \frac12\ln\frac{60}{50}
\simeq 9.1\times10^{-2}.
\label{eq:CMBlogwindow}
\end{equation}}
{The observable interval is therefore short in logarithmic field distance. This gives an operational bound rather than only a formal criterion. Requiring the running correction to change the local quartic by less than a fraction $\varepsilon_{\rm RG}$ over the CMB window gives}
{
\begin{equation}
|\beta_\lambda|
\lesssim
\frac{\varepsilon_{\rm RG}|\lambda_{\rm eff}|}{\Delta\ln\varphi_{50-60}}
\simeq
1.1\times10^{-3}
\left(\frac{\varepsilon_{\rm RG}}{10^{-2}}\right)
\left(\frac{|\lambda_{\rm eff}|}{10^{-2}}\right),
\label{eq:betalambdabound}
\end{equation}}
{with the same replacement $\lambda_{\rm eff}\to\xi$ for the fractional running of the nonminimal coupling. A generic weakly coupled scalar threshold gives a quartic beta function of order}
{
\begin{equation}
\beta_\lambda^{\rm loop}
\sim c_\lambda\frac{\lambda_{\rm eff}^2}{16\pi^2}
\simeq
6.3\times10^{-7}\,c_\lambda
\left(\frac{\lambda_{\rm eff}}{10^{-2}}\right)^2,
\label{eq:loopbetalambdaestimate}
\end{equation}}
{where $c_\lambda$ is a model-dependent coefficient fixed by the fields coupled to the trace-channel scalar \cite{ColemanWeinberg:1973,Machacek:1984zw}. Even for $c_\lambda=O(1$--$10)$, Eq.~\eqref{eq:loopbetalambdaestimate} is well below the one-percent bound in Eq.~\eqref{eq:betalambdabound}. If the hidden-sector matching induces an RG contribution of order the MS coefficient, $\beta_\lambda=O(A)$, then in the controlled-deformation region used for the benchmark scans, $A/\lambda\lesssim10^{-4}$--$10^{-3}$, the integrated shift is again parametrically below $\lambda_{\rm eff}$. The fixed-coupling scan should therefore be read as a pivot-scale EFT approximation with an explicit acceptance criterion, not as the statement that $\beta_\lambda$ and $\beta_\xi$ vanish. These inequalities are technically natural in a weakly coupled gravitational EFT or in a matching scheme where the residual running is higher order over the CMB window, consistent with the standard treatment of scalar couplings in curved-space EFTs \cite{Callan:1970,BirrellDavies:1982,Buchbinder:1992,ParkerToms:2009}. They are also the same kind of finite-field-window assumption used when comparing nonminimal plateau models with CMB observables after choosing a matching prescription \cite{Barvinsky:2008ia,Hertzberg:2010dc}. If a specified hidden sector violates Eq.~\eqref{eq:betalambdabound} or the corresponding $\xi$ bound, the appropriate analysis is the RG-improved version of Eq.~\eqref{eq:VJfull}; the anomaly coefficient $A$ itself is still not identified with $\beta_\lambda$ but fixed by nonperturbative trace-channel matching.}

\textbf{\textit{{Weyl transformation and the exact Einstein-frame action}}}

To compute inflationary observables we go to the Einstein frame with metric
\begin{equation}
g_{\mu\nu}^{E}=\Omega^2(\varphi)\,g_{\mu\nu},
\qquad
\Omega^2(\varphi)=\frac{F(\varphi)}{M_{\rm Pl}^2}.
\label{eq:Weyl}
\end{equation}
Standard Weyl-transformation identities yield
\begin{equation}
S_E=\int d^4x\sqrt{-g_E}\left[
\frac{M_{\rm Pl}^2}{2}R_E
-\frac12K(\varphi)(\nabla_E\varphi)^2
-U(\varphi)
\right],
\label{eq:EinsteinAction}
\end{equation}
with Einstein-frame potential
\begin{equation}
U(\varphi)=\frac{V_J(\varphi)}{\Omega^4}
=\frac{M_{\rm Pl}^4\,V_J(\varphi)}{F(\varphi)^2},
\label{eq:Udef}
\end{equation}
and a nontrivial field-space metric (kinetic prefactor)
\begin{equation}
K(\varphi)=\frac{M_{\rm Pl}^2}{F(\varphi)}
+\frac{3M_{\rm Pl}^2}{2}\left(\frac{F'(\varphi)}{F(\varphi)}\right)^2,
\qquad
F'(\varphi)=2\xi\varphi.
\label{eq:Kdef}
\end{equation}
The canonically normalized Einstein-frame scalar \(\phi\) is defined by
\begin{equation}
\left(\frac{d\phi}{d\varphi}\right)^2 = K(\varphi).
\label{eq:canonNorm}
\end{equation}
Equations \eqref{eq:Udef}--\eqref{eq:canonNorm} are exact and will be the basis of both the analytic slow-roll expansion and the numerical verification.

\textbf{\textit{{Large-\(\xi\) asymptotics and the plateau structure}}}
The inflationary regime of interest is typically
\begin{equation}
\xi \gg 1,\qquad \xi\varphi^2 \gg M_{\rm Pl}^2,
\label{eq:inflationRegime}
\end{equation}
for which
\begin{equation}
F(\varphi)\simeq \xi\varphi^2,\qquad
\Omega^2\simeq \frac{\xi\varphi^2}{M_{\rm Pl}^2}.
\end{equation}
In this limit the second term in \eqref{eq:Kdef} dominates, giving
\begin{equation}
K(\varphi)\simeq \frac{6M_{\rm Pl}^2}{\varphi^2},
\qquad\Rightarrow\qquad
\phi \simeq \sqrt{6}\,M_{\rm Pl}\,\ln\!\left(\frac{\sqrt{\xi}\,\varphi}{M_{\rm Pl}}\right)+{\rm const},
\label{eq:phiLog}
\end{equation}
which is the logarithmic stretching that produces slow roll.
Moreover, because \(V_J(\varphi)\sim \varphi^4\) at large field, the Einstein-frame potential approaches a constant:
\begin{align}
U(\varphi)
&=
\frac{M_{\rm Pl}^4}{\xi^2}\left[
\frac{\lambda}{4} + A\left(\ln\frac{\varphi}{\mu}-\frac14\right)
+ \mathcal{O}\!\left(\frac{M_{\rm Pl}^2}{\xi\varphi^2}\right)
\right].
\label{eq:UlargeExact}
\end{align}

{Equation~\eqref{eq:phiLog} gives $\ln(\varphi/\mu)=\phi/(\sqrt{6}M_{\rm Pl})+{\rm const}$. The MS term therefore appears in the canonical plateau as a slowly varying logarithmic tilt. The Starobinsky-like attractor is recovered when this tilt is subdominant over the observable window, quantified below by $\Delta_{\rm MS}(\varphi_*)\ll1$. Thus the anomaly term does not destroy the plateau; it supplies a controlled deformation of the plateau height, with the same overall $1/\xi^2$ suppression as the quartic contribution.}

% ==========================================================
\section{Inflationary dynamics and observables}
\label{sec:inflation}
% ==========================================================
Although the dynamics is simplest in terms of the canonical field \(\phi\), for analytic manipulation and especially for numerical work it is extremely convenient to express slow-roll quantities directly in terms of the Jordan variable \(\varphi\) using the field-space metric \(K(\varphi)\). Define derivatives with respect to \(\varphi\) by primes. Then
\begin{equation}
\frac{dU}{d\phi}=\frac{U'}{\sqrt{K}},
\qquad
\frac{d^2U}{d\phi^2}=\frac{1}{K}\left(U''-\frac{K'}{2K}U'\right).
\label{eq:dUdphi}
\end{equation}
The exact slow-roll parameters are therefore
\begin{equation}
\epsilon(\varphi)=\frac{M_{\rm Pl}^2}{2}\left(\frac{U_{,\phi}}{U}\right)^2
=
\frac{M_{\rm Pl}^2}{2K(\varphi)}\left(\frac{U'(\varphi)}{U(\varphi)}\right)^2,
\label{eq:epsExact}
\end{equation}
\begin{equation}
\eta(\varphi)=M_{\rm Pl}^2\frac{U_{,\phi\phi}}{U}
=
\frac{M_{\rm Pl}^2}{K(\varphi)\,U(\varphi)}
\left(
U''(\varphi)-\frac{K'(\varphi)}{2K(\varphi)}U'(\varphi)
\right).
\label{eq:etaExact}
\end{equation}
Inflation ends when
\begin{equation}
\epsilon(\varphi_{\rm end})=1.
\label{eq:end}
\end{equation}
The number of e-folds from \(\varphi_*\) to \(\varphi_{\rm end}\) is
\begin{equation}
N
\simeq
\frac{1}{M_{\rm Pl}^2}\int_{\phi_{\rm end}}^{\phi_*}\frac{U}{U_{,\phi}}\,d\phi
=
\frac{1}{M_{\rm Pl}^2}\int_{\varphi_{\rm end}}^{\varphi_*}
\frac{U(\varphi)\,K(\varphi)}{U'(\varphi)}\,d\varphi,
\label{eq:NExact}
\end{equation}
where we used \(d\phi=\sqrt{K}\,d\varphi\) and \eqref{eq:dUdphi}.
Finally, the leading CMB observables at horizon exit are
\begin{equation}
A_s \simeq \frac{U_*}{24\pi^2M_{\rm Pl}^4\epsilon_*},
\quad
n_s \simeq 1-6\epsilon_*+2\eta_*,
\quad
r\simeq 16\epsilon_*,
\label{eq:ObsExact}
\end{equation}
with all starred quantities evaluated at \(\varphi=\varphi_*\).
Equations \eqref{eq:epsExact}--\eqref{eq:ObsExact} provide a complete and internally consistent set of formulae that can be evaluated analytically in limiting regimes and numerically without ambiguity.

{
\textbf{\textit{{Pure MS baseline before the gravitational completion}}}

Before turning on the two gravitational EFT coefficients, it is instructive to set
}
{
\begin{equation}
\lambda=0,\qquad \xi=0,
\label{eq:pureMSlimit}
\end{equation}}

{and to uplift the MS vacuum by adding $A\mu^4/4$, so that}
{
\begin{equation}
V_{\rm pure}(\varphi)=A\varphi^4\left(\ln\frac{\varphi}{\mu}-\frac14\right)+\frac{A\mu^4}{4}.
\label{eq:pureMSuplift}
\end{equation}}

{The overall coefficient $A$ cancels from $n_s$ and $r$ and is fixed only afterward by $A_s$. The shape prediction is therefore controlled by $\mu/M_{\rm Pl}$ and by the branch of the potential. The large-field branch behaves approximately as a logarithmically corrected quartic potential and gives tensors of order $r\gtrsim 0.1$ for $N=55$ over the scanned range. The small-field branch can suppress $r$, but the observed tilt is reached only for trans-Planckian $\mu$ and typically trans-Planckian horizon-exit field values. This does not mean that the pure MS theory is mathematically inconsistent; it means that, as a controlled single-field inflationary EFT, it becomes sensitive to the same higher-dimensional operators and UV data that any trans-Planckian scalar excursion probes. The pure MS scalar is therefore an essential theoretical baseline, but not the controlled regime used for the CMB-compatible model.}

\begin{figure}[t]
\centering
\includegraphics[width=0.92\linewidth]{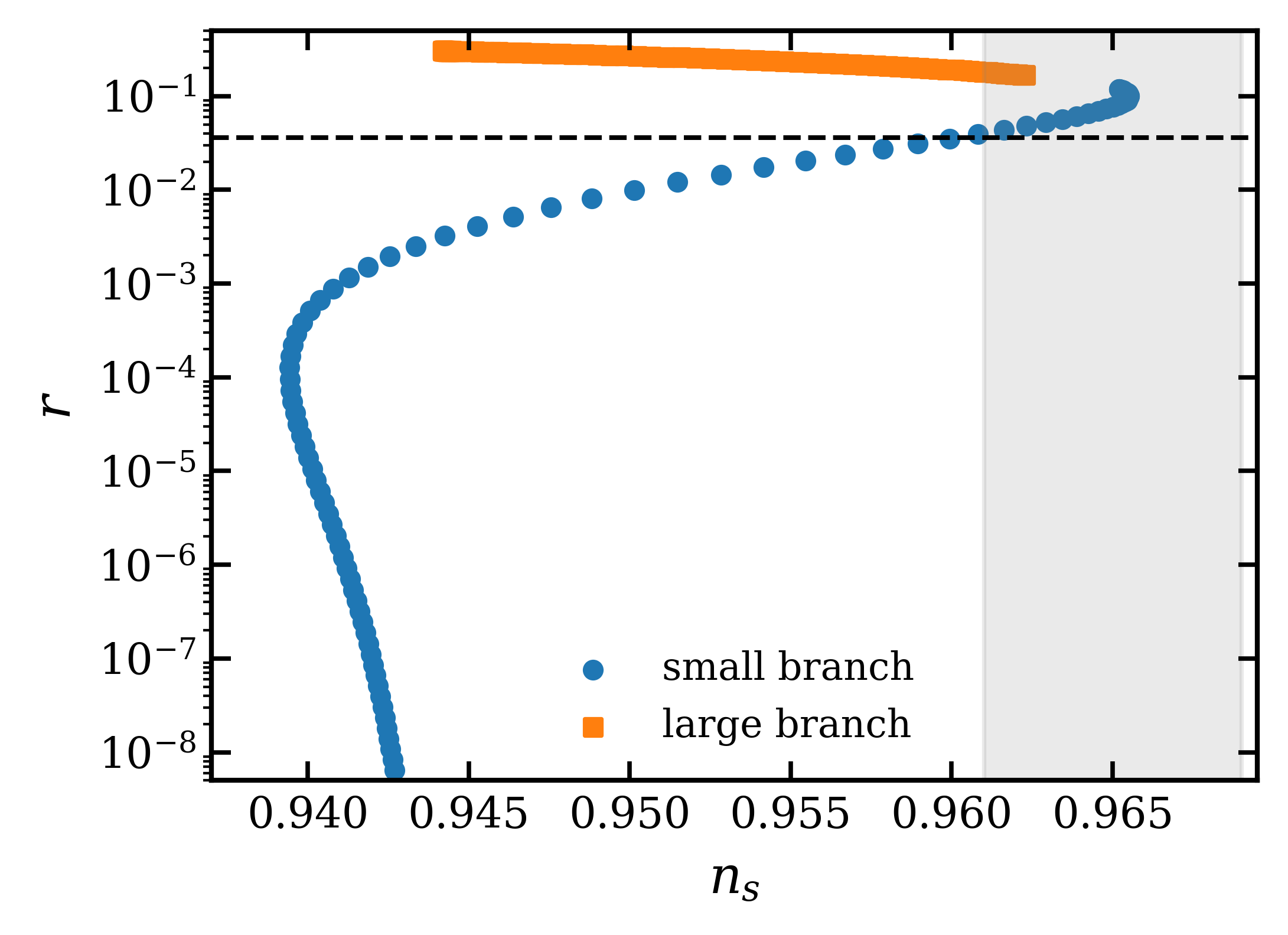}
\caption{{Pure MS minimal predictions at $N=55$ with $\lambda=\xi=0$, computed from Eq.~\eqref{eq:pureMSuplift}. The large-field branch is tensor dominated, while the small-field branch reaches the preferred tilt only in a trans-Planckian matching-scale regime. The horizontal dashed line marks $r=0.036$.}}
\label{fig:pureMSnsr}
\end{figure}

\begin{figure}[t]
\centering
\includegraphics[width=0.92\linewidth]{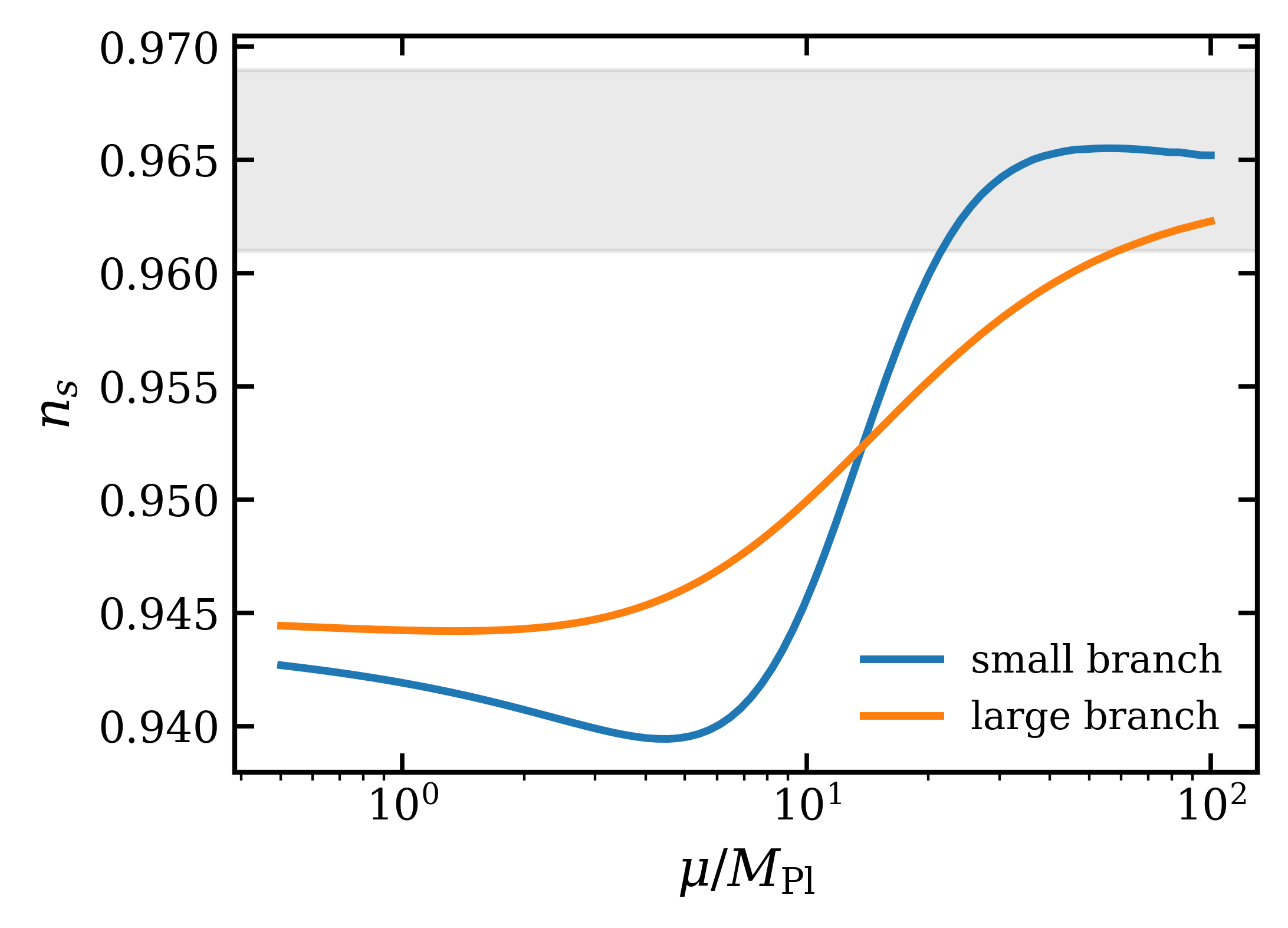}
\caption{{Scalar tilt in the pure MS minimal limit as a function of $\mu/M_{\rm Pl}$ at $N=55$. The gray band indicates the observationally preferred region. The plot shows why the pure anomaly-matched scalar, without the nonminimal flattening, is not the robust controlled inflationary regime.}}
\label{fig:pureMSmu}
\end{figure}

\begin{table}[t]
\centering
\begin{tabular}{ccccc}
\toprule
branch & $\mu/M_{\rm Pl}$ & $\varphi_*/M_{\rm Pl}$ & $n_s$ & $r$\\
\midrule
small & $10$ & $2.10$ & $0.9456$ & $4.42\times10^{-3}$\\
small & $20$ & $8.78$ & $0.9599$ & $3.51\times10^{-2}$\\
small & $25$ & $12.99$ & $0.9628$ & $5.10\times10^{-2}$\\
large & $10$ & $28.50$ & $0.9499$ & $2.69\times10^{-1}$\\
large & $50$ & $66.01$ & $0.9604$ & $1.87\times10^{-1}$\\
\bottomrule
\end{tabular}
\caption{{Representative pure-MS minimal results for $N=55$. Since $A$ cancels out of the shape observables, these values isolate the prediction before the gravitational Wilson coefficients $\lambda$ and $\xi$ are added.}}
\label{tab:pureMSbaseline}
\end{table}

\textbf{\textit{{Attractor limit and analytic predictions}}}

We now show explicitly how the standard attractor predictions emerge in the regime \eqref{eq:inflationRegime} when the plateau is controlled mainly by the quartic term.
For clarity, first set \(A=0\) (pure nonminimal quartic), for which
\begin{equation}
V_J(\varphi)=\frac{\lambda}{4}\varphi^4,\qquad
U(\varphi)=\frac{\lambda\,M_{\rm Pl}^4}{4}\frac{\varphi^4}{\left(M_{\rm Pl}^2+\xi\varphi^2\right)^2}.
\end{equation}
Introduce the convenient variable
\begin{equation}
y(\varphi)\equiv \frac{\xi\varphi^2}{M_{\rm Pl}^2},
\qquad\Rightarrow\qquad
\Omega^2=1+y.
\end{equation}
Then
\begin{equation}
U(\varphi)=\frac{\lambda M_{\rm Pl}^4}{4\xi^2}\left(\frac{y}{1+y}\right)^2
=\frac{\lambda M_{\rm Pl}^4}{4\xi^2}\left(1-\frac{1}{1+y}\right)^2.
\end{equation}
In the large-field regime \(y\gg 1\), the quantity \((1+y)^{-1}\) is exponentially small in the canonical field. Importantly, the mapping between $y$ and the canonical field is controlled by the kinetic prefactor $K(\varphi)$, which depends only on $F(\varphi)$ and is therefore independent of the detailed form of $V_J(\varphi)$. Indeed, from \eqref{eq:Kdef} one finds in this regime
\begin{equation}
\frac{d\phi}{d\varphi}\simeq \sqrt{\frac{6}{\varphi^2}}\,M_{\rm Pl}
\quad\Rightarrow\quad
\phi\simeq \sqrt{\frac{3}{2}}\,M_{\rm Pl}\,\ln(1+y) + \text{const},
\end{equation}
which implies the standard relation
\begin{equation}
1+y \;\simeq\; e^{\sqrt{\frac{2}{3}}\frac{\phi}{M_{\rm Pl}}}.
\label{eq:yphi}
\end{equation}
Substituting \eqref{eq:yphi} gives the familiar plateau form
\begin{equation}
U(\phi)=\frac{\lambda M_{\rm Pl}^4}{4\xi^2}\left(1-e^{-\sqrt{\frac{2}{3}}\frac{\phi}{M_{\rm Pl}}}\right)^2,
\label{eq:plateau}
\end{equation}
which is the same functional form as Starobinsky inflation.
From \eqref{eq:plateau} one obtains, for large \(\phi\),
\begin{equation}
\epsilon(\phi)\simeq \frac{4}{3}\,e^{-2\sqrt{\frac{2}{3}}\frac{\phi}{M_{\rm Pl}}},
\qquad
N\simeq \frac{3}{4}\,e^{\sqrt{\frac{2}{3}}\frac{\phi_*}{M_{\rm Pl}}},
\end{equation}
and therefore at leading order in \(1/N\),
\begin{equation}
\epsilon_* \simeq \frac{3}{4N^2},
\qquad
n_s \simeq 1-\frac{2}{N},
\qquad
r\simeq \frac{12}{N^2}.
\label{eq:attractorPred}
\end{equation}
This is the attractor behavior shared by a broad class of nonminimal plateau models \cite{Salopek:1989,Fakir:1990,Bezrukov:2008,Kaiser:2010ps,Hertzberg:2010dc,Kallosh:2014}.

\textbf{\textit{{Including the MS anomaly term as a logarithmic deformation}}}

We now restore the MS term with \(A\neq 0\).
At large field, using \eqref{eq:UlargeExact},
\begin{equation}
U(\varphi)\simeq \frac{M_{\rm Pl}^4}{\xi^2}\left[\frac{\lambda}{4}
+ A\left(\ln\frac{\varphi}{\mu}-\frac14\right)\right].
\end{equation}
The key point is that the plateau is preserved; the MS term modifies the plateau height and introduces a mild \(\ln\varphi\) dependence. {Consequently, the MS sector affects the observables mainly through the local slope and curvature of the plateau across the limited CMB interval, not through a new asymptotic large-field behavior. This is why the deformation can be small in $n_s$ and $r$ while still carrying microscopic information: its coefficient and matching scale are not arbitrary fit functions but trace-channel data of the hidden confining theory.}

For analytic control, it is useful to delineate the regime in which the MS term acts as a \emph{perturbative} deformation of the quartic plateau over the observational window. A convenient calculable quantity is
\begin{equation}
\Delta_{\rm MS}(\varphi)\;\equiv\;\left|\frac{A}{\lambda}\ln\frac{\varphi}{\mu}\right|.
\end{equation}
and we demand:
\begin{equation}
\left|\frac{A}{\lambda}\ln\frac{\varphi_*}{\mu}\right|\ll 1,
\label{eq:deformSmall}
\end{equation}
When $\Delta_{\rm MS}(\varphi_*)\ll 1$ at horizon exit, the leading predictions \eqref{eq:attractorPred} remain intact with calculable corrections. We emphasize that this is not a fundamental consistency condition of the model: it simply identifies the domain where the attractor approximation and small-deformation analytic expansions are accurate. Our numerical analysis (Sec.~\ref{numerics}) uses the exact slow-roll expressions \eqref{eq:epsExact}--\eqref{eq:ObsExact} and remains valid also when $\Delta_{\rm MS}$ is not parametrically small.
In addition to $\Delta_{\rm MS}$, a useful measure of how strongly the MS term
affects the slow-roll slope on the plateau is $A/\mathcal B_*$ evaluated at horizon exit.
In the CMB-normalized benchmark with $\mu=10^{16}\,{\rm GeV}$ and $\alpha\in[0.01,0.03]$ we find
$\Delta_{\rm MS}(\varphi_*)\simeq 0.04\text{--}0.12$ and $A/\mathcal B_*\simeq 0.035\text{--}0.082$.
This provides a clean theoretical interpretation: the MS term is fixed by anomaly matching in the trace-channel EFT; inflation is made viable by the nonminimal coupling which converts quartic growth into an Einstein-frame plateau; and the MS term yields a theoretically motivated logarithmic imprint whose size is governed by $A/\lambda$ and can be quantified by $\Delta_{\rm MS}$.

\textbf{\textit{{Scalar amplitude normalization and parameter relations}}}

The observed scalar amplitude fixes the overall height of the plateau.
In the attractor regime (dominantly quartic plateau), one may use
\begin{equation}
U_* \simeq \frac{\lambda M_{\rm Pl}^4}{4\xi^2},
\qquad
\epsilon_*\simeq \frac{3}{4N^2}.
\end{equation}
Substituting into \eqref{eq:ObsExact} gives
\begin{equation}
A_s \simeq \frac{\lambda N^2}{72\pi^2\xi^2},
\label{eq:AsRel}
\end{equation}
hence
\begin{equation}
\xi \simeq \frac{N}{\sqrt{72}\,\pi}\,\sqrt{\frac{\lambda}{A_s}}
\;\simeq\;
4.5\times 10^4\,\sqrt{\lambda}\,
\left(\frac{N}{55}\right)\left(\frac{2.1\times 10^{-9}}{A_s}\right)^{1/2},
\label{eq:xiLamRel}
\end{equation}
consistent with the familiar scaling from Higgs-inflation-like models \cite{Bezrukov:2008}.
The MS parameters \(A\) and \(\mu\) are tied to the underlying confining sector through \eqref{eq:varphiDef}, \eqref{eq:AMap}. Condition \eqref{eq:deformSmall} then translates into an explicit inequality on \((m,|e_{\rm vac}|)\) relative to \(\lambda\), controlling the size of anomaly-induced deformations during inflation.

{The amplitude constraint is most transparently expressed in terms of the plateau bracket}
{
\begin{equation}
U_* \simeq \frac{M_{\rm Pl}^4}{\xi^2}\,\mathcal{B}_*,\qquad
\mathcal{B}_*\equiv \left[\frac{\lambda}{4}+A\left(\ln\frac{\varphi_*}{\mu}-\frac14\right)\right].
\label{eq:Bdef}
\end{equation}}
{Combining this expression with $A_s=U_*/(24\pi^2M_{\rm Pl}^4\epsilon_*)$ gives}
{
\begin{equation}
\mathcal B_* =24\pi^2 A_s\, \epsilon_*\, \xi^2,
\label{eq:Bexact}
\end{equation}}
{which is exact at leading slow-roll order once the numerically determined $\epsilon_*$ is used. In the attractor approximation $\epsilon_*\simeq3/(4N^2)$, this reduces to}
{
\begin{equation}
A_s \simeq \frac{\mathcal{B}_*}{\xi^2}\,\frac{N^2}{18\pi^2},\qquad
\mathcal{B}_* \simeq 18\pi^2 A_s\frac{\xi^2}{N^2}.
\label{eq:AsBrel}
\end{equation}}
{Thus the relevant energy density is the Einstein-frame plateau height, not the Jordan-frame quartic evaluated without the Weyl suppression. Horizon exit occurs at $y_*=\xi\varphi_*^2/M_{\rm Pl}^2=O(N)$; in the quartic attractor $y_*\simeq4N/3$, so $\Omega_*^4=(1+y_*)^2$ supplies the familiar large suppression. This is why $\mathcal B_*$ can be small for $\xi\sim10^3$ or order one for $\xi\sim10^4$--$10^5$ without conflicting with $A_s$.}

{The MS term is phenomenologically relevant when it changes $\mathcal B_*$ or its slope across the CMB interval. Two regimes are worth separating. In the comparable-contribution regime, $A\ln(\varphi_*/\mu)$ is of the same order as $\lambda/4$ and the logarithm substantially reshapes the plateau. In the controlled-deformation regime, $|A/\lambda|\ll1$ and the logarithm gives a small but calculable anomaly-matched correction. The numerical scan below covers both possibilities and identifies the subset compatible with the observed tilt, tensor bound, and EFT hierarchy.}

Our scan figures are designed to identify which regime is realized for given $(\xi,\lambda,A,\mu)$ and to demonstrate that viable points exist.

\section{Numerical Scan}\label{numerics}

The analytic results above are complemented by a direct numerical evaluation of the exact slow-roll expressions \eqref{eq:epsExact}--\eqref{eq:ObsExact} with the full potential \eqref{eq:VJfull}.
The numerical procedure is unambiguous:
\begin{enumerate}
\item For given \((\xi,\lambda,A,\mu)\), construct \(F(\varphi)\), \(U(\varphi)\), and \(K(\varphi)\) from \eqref{eq:Udef}--\eqref{eq:Kdef}.
\item Compute \(\epsilon(\varphi)\) and solve \(\epsilon(\varphi_{\rm end})=1\) for \(\varphi_{\rm end}\).
\item Compute \(N(\varphi)\) via the exact integral \eqref{eq:NExact} and invert to obtain \(\varphi_*\) for the desired \(N\).
\item Evaluate \(n_s,r,A_s\) at \(\varphi_*\) using \eqref{eq:ObsExact}, and optionally compute \(\alpha_s\equiv dn_s/d\ln k\simeq -dn_s/dN\) by finite differencing in \(N\).
\end{enumerate}
This algorithm is exactly what our numerical code implements and is the appropriate way to validate the model because it does not rely on asymptotic canonical-field expressions: all nontrivial field-space factors are included through \(K(\varphi)\).

\vspace{0.5em}
\noindent
Two conceptual points motivate this numerical strategy. First, it provides a strict internal consistency check on the analytic treatment: the same Einstein-frame potential \(U(\varphi)\) and field-space metric \(K(\varphi)\) that define the theory in \eqref{eq:EinsteinAction} are used directly in \eqref{eq:ObsExact}. In particular, the field redefinition \eqref{eq:canonNorm} is never approximated; its effects enter exactly through \(K(\varphi)\) in \eqref{eq:epsExact}--\eqref{eq:etaExact} and through the e-fold integral \eqref{eq:NExact}. This guarantees that any departure from the attractor predictions arises from physical deformations of the potential (not from a choice of parametrization or an asymptotic truncation).

\noindent
Second, the procedure makes transparent how the MS anomaly term acts in the inflationary regime. Varying \((A,\mu)\) at fixed \((\xi,\lambda)\) modifies \(U(\varphi)\) through a logarithmic deformation while leaving the flattening mechanism controlled by \(F(\varphi)\) intact. The numerical inversion of \eqref{eq:NExact} then directly determines how this deformation shifts the horizon-exit point \(\varphi_*\), and hence the values of \(\epsilon_*\) and \(\eta_*\) that enter \eqref{eq:ObsExact}. In this way, the scan quantitatively identifies the regime in which $\Delta_{\rm MS}(\varphi_*)$ is small (so that the deformation is perturbative) and the regime in which the logarithmic term significantly reshapes the slow-roll trajectory.

\begin{figure}[t]
\centering
\includegraphics[width=0.92\linewidth]{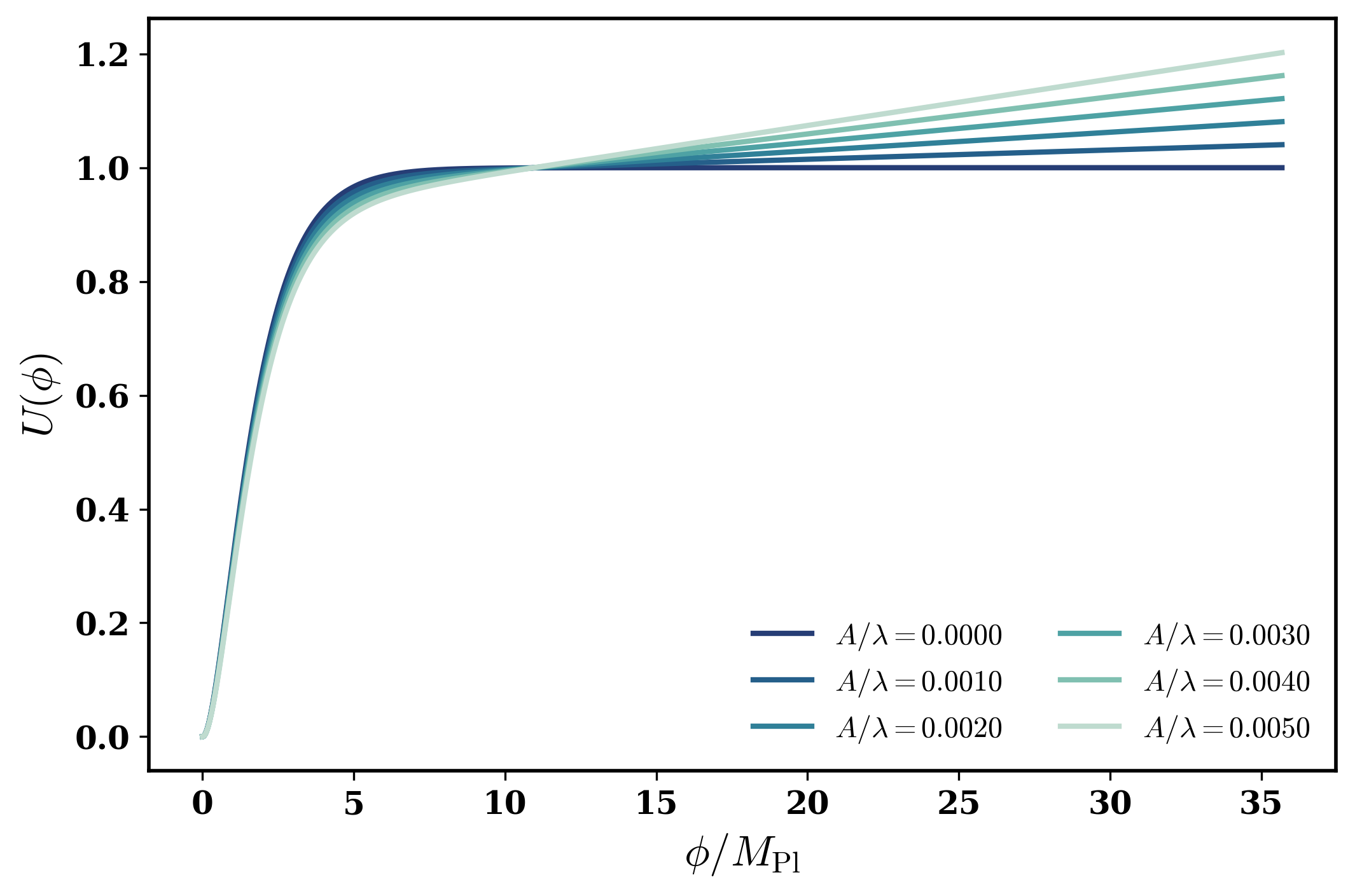}
\caption{Einstein-frame potential \(U(\phi)\) and plateau deformation. The nonminimal coupling produces an asymptotic plateau at large field, while the MS anomaly term induces a mild logarithmic deformation consistent with Eq.~\eqref{eq:UlargeExact} \cite{Callan:1970,Bezrukov:2008,Kallosh:2014}.}
\label{fig:Uphi}
\end{figure}

\begin{figure}[t]
\centering
\includegraphics[width=0.92\linewidth]{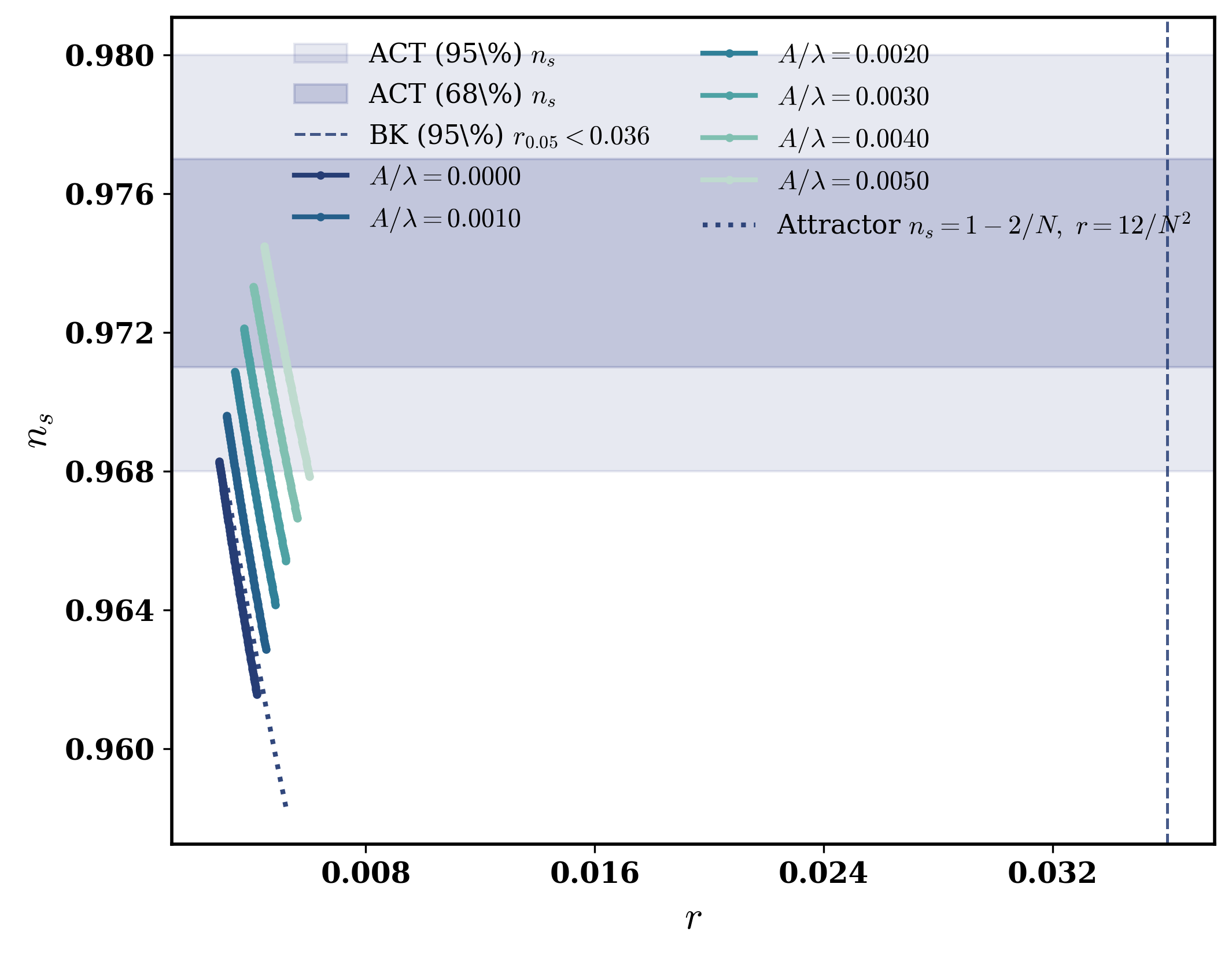}
\caption{Numerical predictions in the \((n_s,r)\) plane. The scan evaluates \(n_s\) and \(r\) using Eqs.~\eqref{eq:epsExact}--\eqref{eq:ObsExact}. The attractor curve emerges at large \(\xi\) when Eq.~\eqref{eq:deformSmall} holds \cite{Bezrukov:2008,Kallosh:2014}.}
\label{fig:nsr}
\end{figure}

\begin{figure}[t]
\centering

\includegraphics[width=0.92\linewidth]{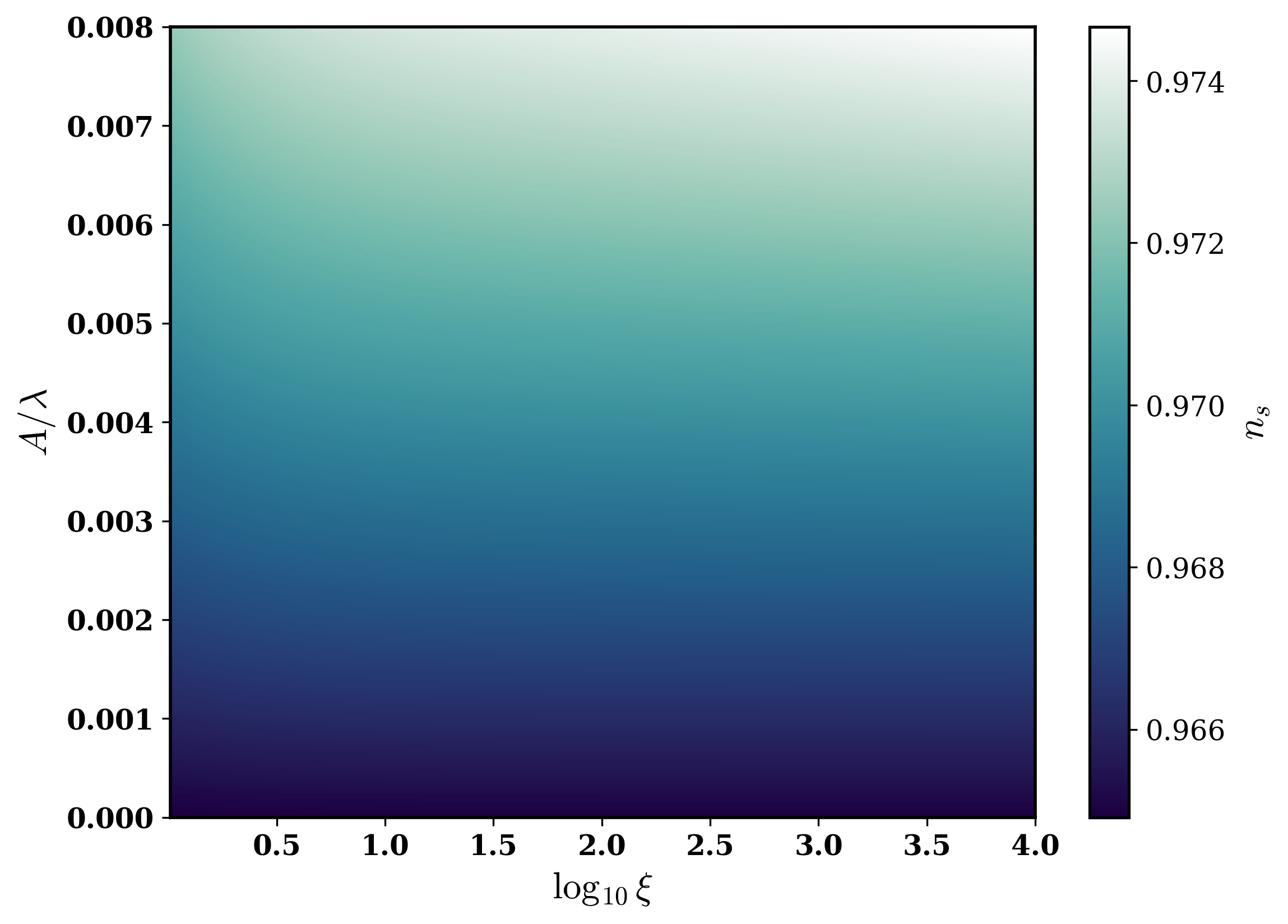}
\caption{Heatmap of \(n_s\) over parameter space. Scan in \((\log_{10}\xi, A/\lambda)\) illustrating the approach to the attractor at large \(\xi\) and controlled departures as \(A/\lambda\) increases.}
\label{fig:heat_ns}
\end{figure}

\begin{figure}[t]
\centering

\includegraphics[width=0.92\linewidth]{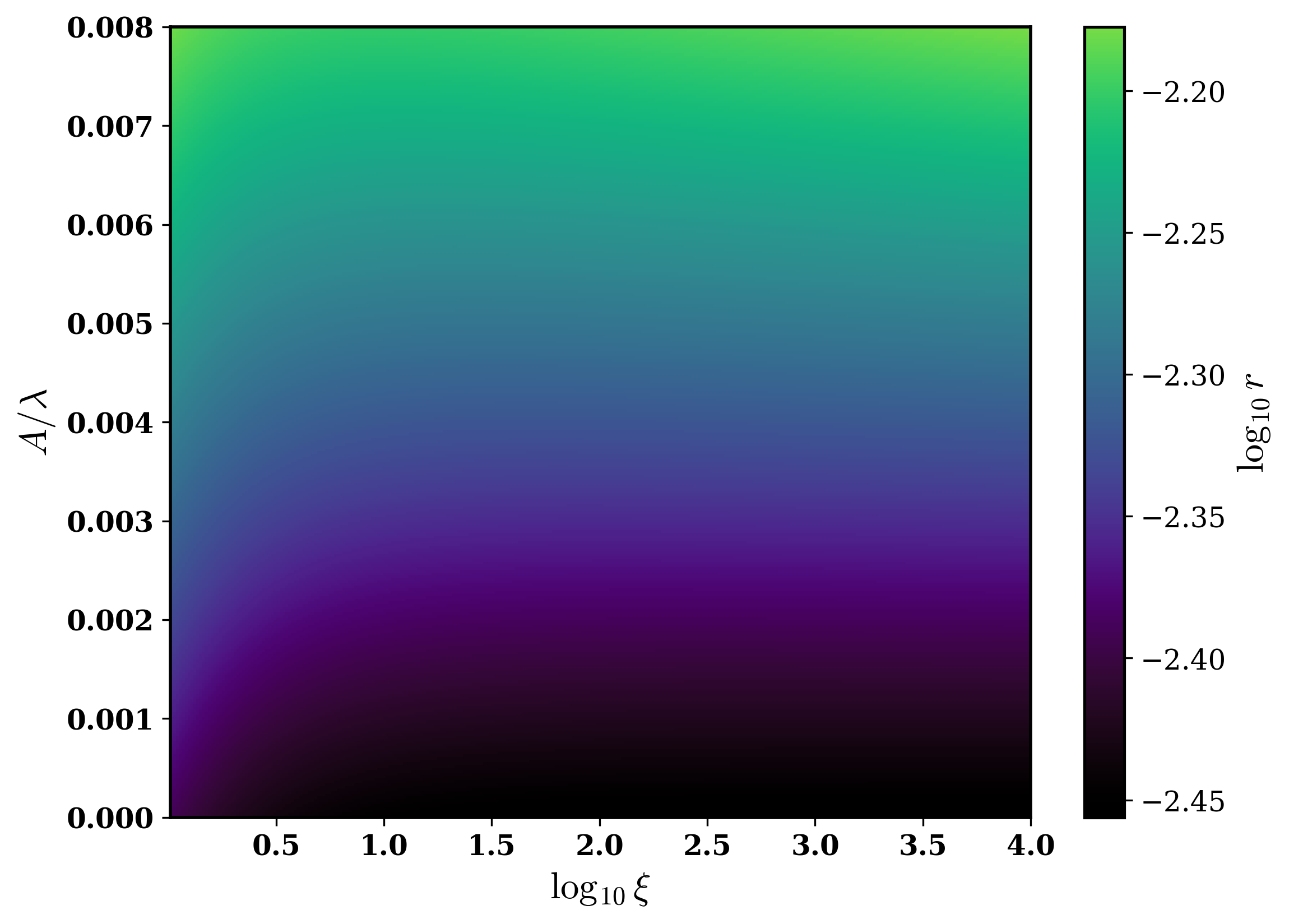}
\caption{Heatmap of \(\log_{10} r\) over parameter space. The tensor amplitude decreases toward the plateau regime, consistent with the scaling \(r\simeq 12/N^2\) in the attractor domain \cite{Kallosh:2014}.}
\label{fig:heat_r}
\end{figure}

\begin{figure*}[t]
\centering
\subfloat[$n_s(N)$ from exact slow roll.%
\label{fig:app_ns_of_N}]{%
\includegraphics[width=0.48\textwidth]{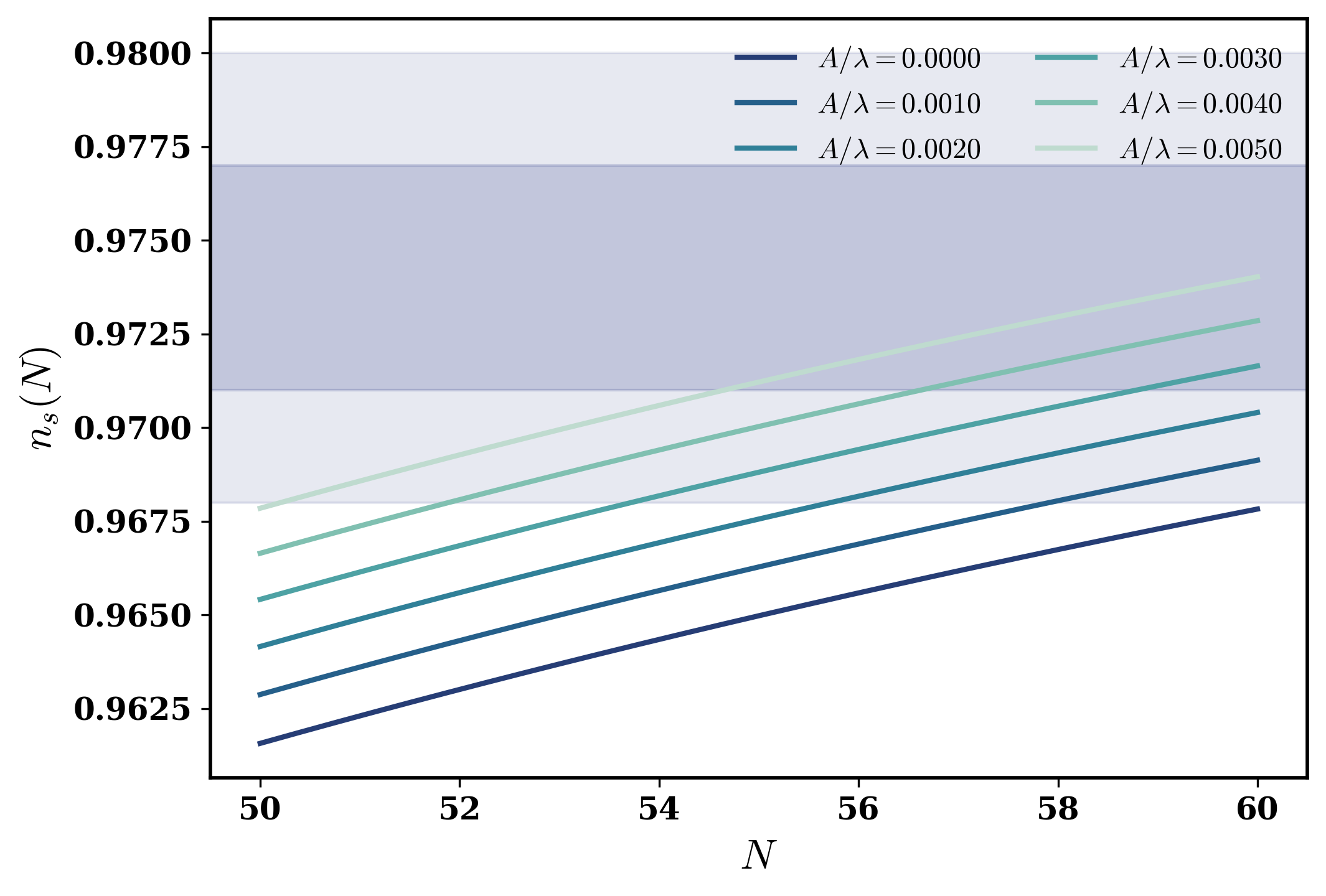}}
\hfill
\subfloat[\textbf{$r(N)=16\epsilon_*(N)$} from exact slow roll.%
\label{fig:app_r_of_N}]{%
\includegraphics[width=0.48\textwidth]{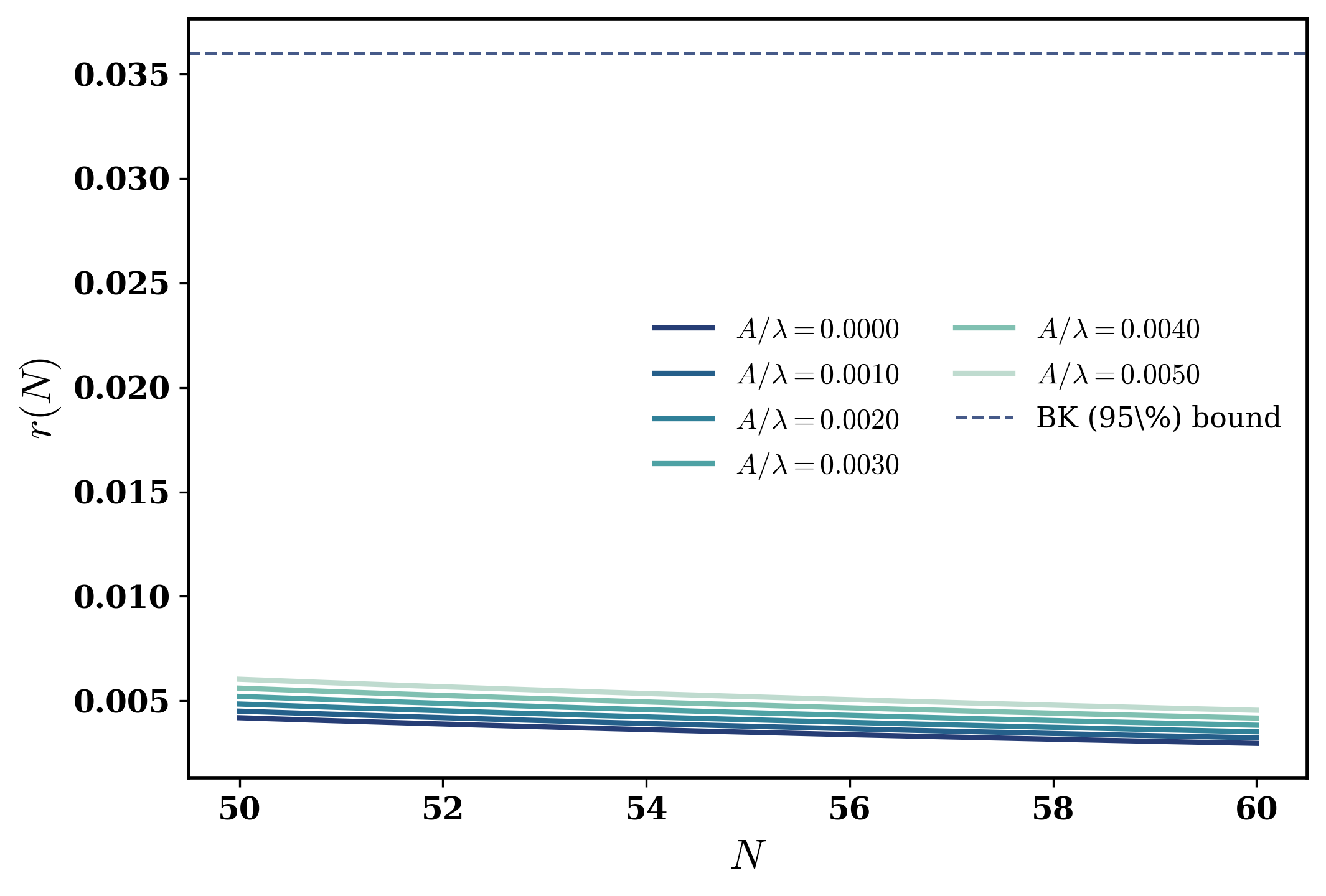}}\\[0.8ex]
\subfloat[$\alpha_s(N)\simeq -dn_s/dN$ (finite differencing).%
\label{fig:app_alpha_s_of_N}]{%
\includegraphics[width=0.48\textwidth]{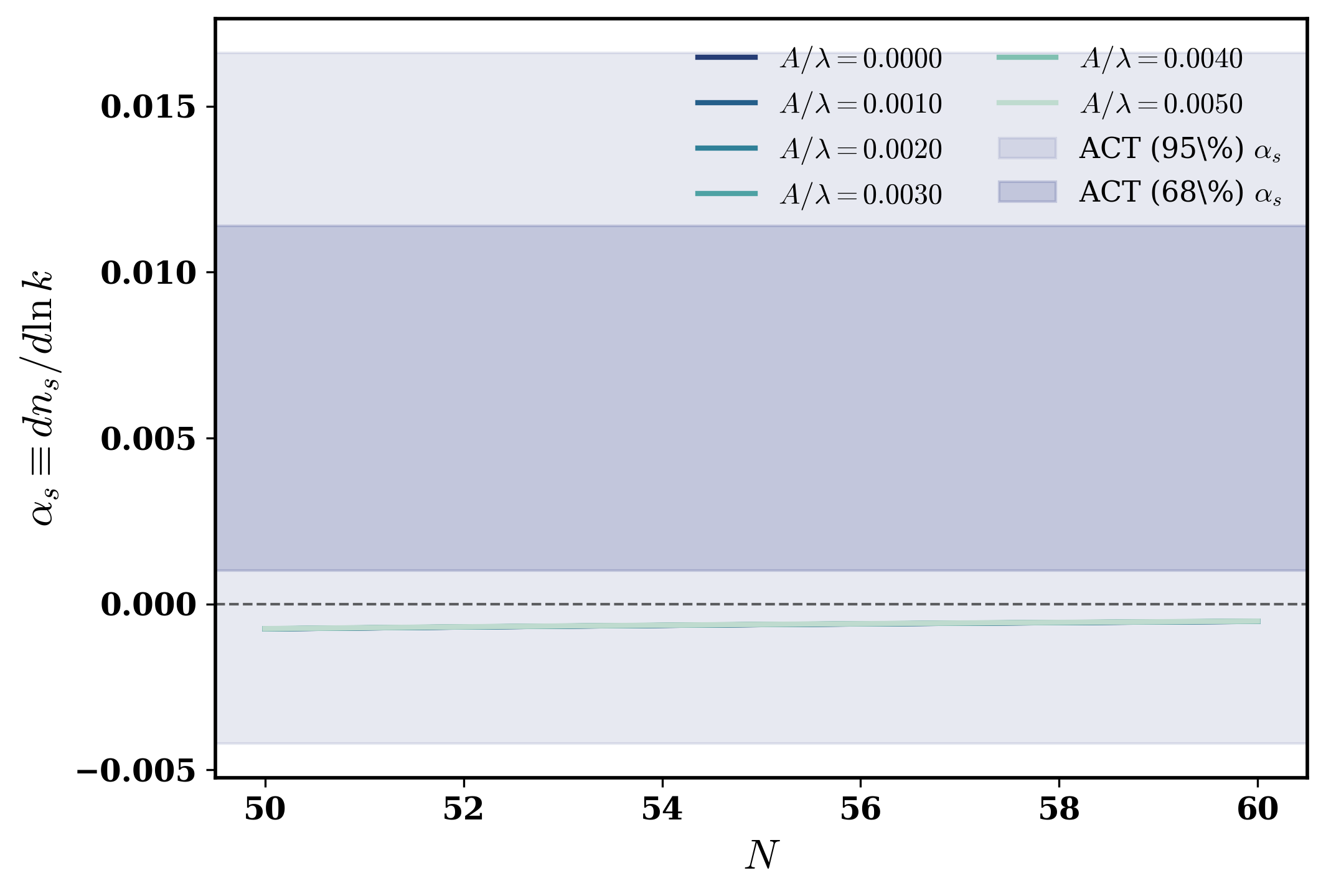}}
\hfill
\subfloat[Heatmap of $\alpha_s$ at fixed $N=55$ in $(\log_{10}\xi,A/\lambda)$.%
\label{fig:app_heatmap_alpha_s}]{%
\includegraphics[width=0.48\textwidth]{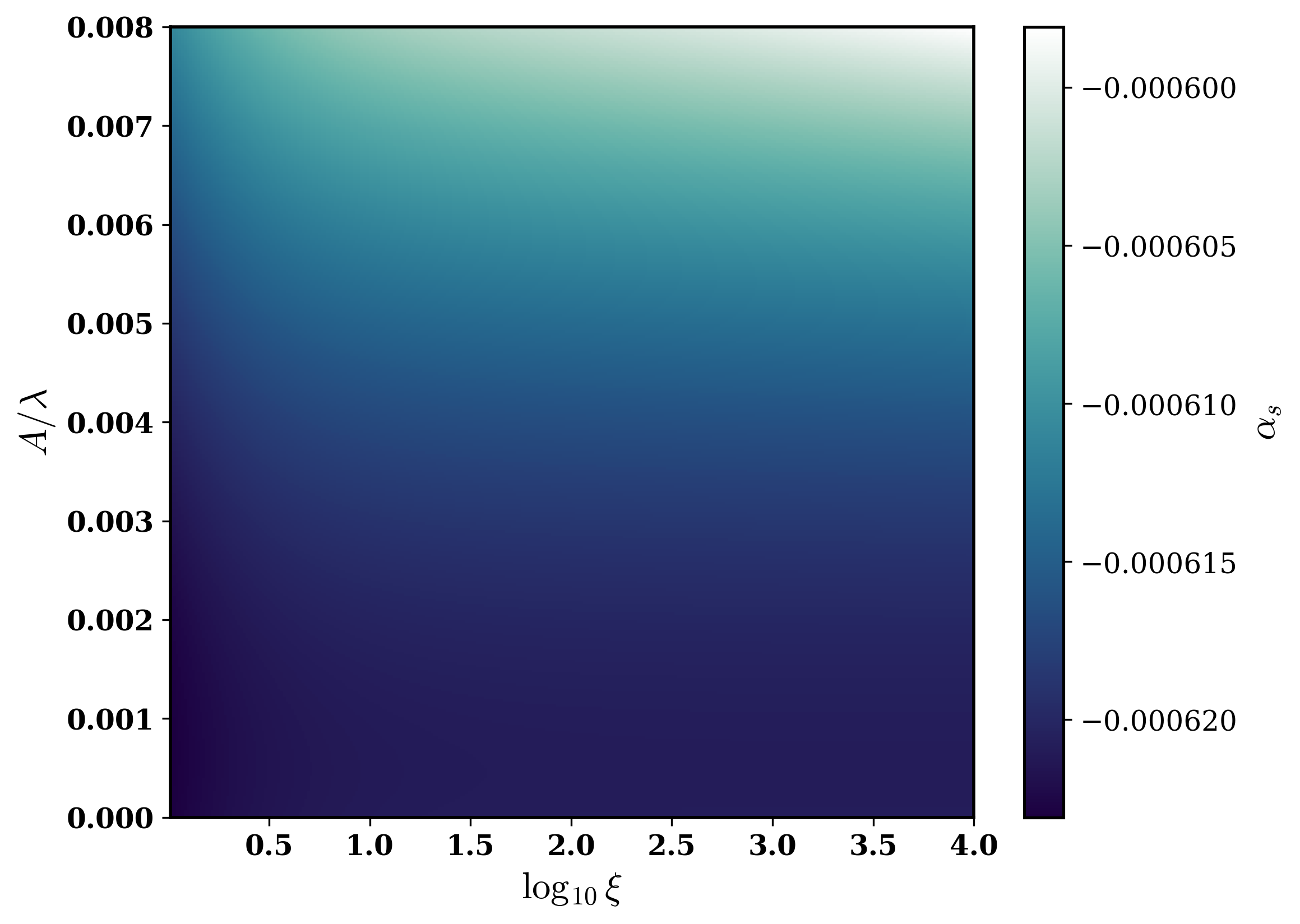}}

\caption{
All quantities are computed using the exact Einstein-frame potential \eqref{eq:Udef}, the exact kinetic prefactor \eqref{eq:Kdef},
and the slow-roll relations \eqref{eq:epsExact}--\eqref{eq:ObsExact}. Panels show: (a) the scalar tilt $n_s$ as a function of $N$;
(b) the tensor-to-scalar ratio $r$ as a function of $N$; (c) the running $\alpha_s\simeq -dn_s/dN$; and
(d) a parameter-space heatmap of $\alpha_s$ at fixed $N=55$.
Together these plots quantify the approach to the strong-coupling attractor and the size of the logarithmic Migdal--Shifman deformation
controlled by $A/\lambda$ and constrained by \eqref{eq:deformSmall}.}
\label{fig:app_fourpanel_diagnostics}
\end{figure*}

{The numerical section is organized to mirror the logic of the EFT. We first show the pure MS baseline, Figs.~\ref{fig:pureMSnsr} and \ref{fig:pureMSmu}, because it fixes what the anomaly-matched scalar predicts before $\lambda$ and $\xi$ are introduced. We then turn on the nonminimal completion and display the plateau observables in Figs.~\ref{fig:Uphi}--\ref{fig:app_fourpanel_diagnostics}. This order is important: the role of $\xi$ is not hidden in the scan but isolated as the physical flattening mechanism required to move the model from the pure-MS baseline to the CMB-compatible plateau regime.}

\subsection{Observational status and numerical interpretation}
\label{subsec:obsstatus}

In the attractor regime the leading predictions \eqref{eq:attractorPred} yield
\begin{equation}
n_s \simeq 1-\frac{2}{N},\qquad r\simeq \frac{12}{N^2},
\end{equation}
which for \(N\simeq 50\)--\(60\) corresponds to \(n_s\simeq 0.96\)--\(0.967\) and \(r\sim \text{few}\times 10^{-3}\). These values lie in the observationally preferred region constrained by current CMB measurements \cite{Planck:2018,BK:2024,ACTDR6LCDM:2025}. The distinctive model-dependent content of our construction is not the existence of the plateau itself---a generic consequence of large nonminimal coupling---but the fact that the leading deformation away from the pure attractor is fixed by anomaly matching \cite{Migdal:1982}. Specifically, the MS term enters as a logarithmic contribution to the Jordan-frame potential and therefore induces a mild, theoretically mandated scale dependence in the Einstein-frame plateau, see \eqref{eq:UlargeExact} \cite{Migdal:1982}. 

Our numerical scans implement the exact slow-roll framework \eqref{eq:epsExact}--\eqref{eq:ObsExact} and therefore quantify, without asymptotic assumptions, how the MS deformation shifts the horizon-exit point \(\varphi_*\) and modifies \((\epsilon_*,\eta_*)\). This provides two complementary outputs: (i) the \((n_s,r)\) locus as \((\xi,\lambda)\) and \(A/\lambda\) are varied at fixed \(N\), and (ii) parameter-plane maps that identify where \eqref{eq:deformSmall} is satisfied and where the deformation becomes phenomenologically relevant. In particular, the scans cleanly separate the universal flattening effect of \(\xi\) (which governs the approach to the attractor) from the anomaly-sector imprint (governed by \(A,\mu\)), thereby giving a controlled interpretation of any deviations from \eqref{eq:attractorPred}. This is the sense in which the MS logarithm is a predictive ingredient: once \((A,\mu)\) are specified by the confining sector, the size and sign of the leading departures from the pure attractor are fixed and can be confronted with CMB constraints, including limits on the running \(\alpha_s\) when included \cite{Planck:2018,ACTDR6Ext:2025,ACTDR6LCDM:2025}.


% ==========================================================
\section{EFT control and confining-sector validity}
\label{sec:discussion}
% ==========================================================

{When promoted to inflation, the central requirement is not only that the nonminimal-gravity sector remain perturbative, but also that the single-field truncation of the confining sector remain below its intrinsic gap. There are therefore two independent scales. The first is the background-dependent strong-coupling scale associated with the nonminimal coupling. The second is the confining-sector cutoff $m_{\rm gap}$, conservatively identified with the mass scale at which heavier glueball resonances enter the trace-channel EFT \cite{Migdal:1982,Morningstar_1999,Lucini:2004my,Lucini:2010nv}. In a Wilsonian inflationary EFT, integrating out heavier states is reliable when their physical frequencies remain above the Hubble scale and above the light adiabatic mass scale, so corrections are suppressed by powers of $H_*^2/m_{\rm gap}^2$ and $m_{\rm eff}^2/m_{\rm gap}^2$ \cite{Cheung:2008,Weinberg:2008,Achucarro:2012sm}.}

{The inflationary background is characterized by}
{
\begin{equation}
H_*^2 \simeq \frac{U_*}{3M_{\rm Pl}^2}.
\end{equation}}

{For the nonminimal coupling, the vacuum estimate $\Lambda_{\rm vac}\sim M_{\rm Pl}/\xi$ is not the relevant scale during inflation. On the large-field background $\xi\varphi^2\gg M_{\rm Pl}^2$, the fluctuation cutoff is parametrically raised; a standard estimate gives}
{
\begin{equation}
\Lambda_{\rm bg}\sim \frac{M_{\rm Pl}}{\sqrt{\xi}},
\end{equation}}

{up to order-one factors \cite{Barbon:2009,Burgess:2009,Giudice:2011}. In the plateau regime $U_*\sim(\lambda/\xi^2)M_{\rm Pl}^4$,}
{
\begin{equation}
\frac{H_*}{\Lambda_{\rm bg}}\sim \sqrt{\frac{\lambda}{\xi}}\ll1,
\label{eq:HoverLambda}
\end{equation}}

{which is easily satisfied on the CMB-normalized locus.}

{The more model-specific constraint is the intrinsic confining-sector one. For a pure $SU(N_h)$ Yang--Mills hidden sector with coupling specified at a UV scale $\mu_{\rm UV}$, the confinement scale is generated by dimensional transmutation,}
{
\begin{equation}
\Lambda_h
=\mu_{\rm UV}\exp\!\left[-\frac{8\pi^2}{b_0g_h^2(\mu_{\rm UV})}\right],
\qquad b_0=\frac{11}{3}N_h,
\label{eq:hiddenLambda}
\end{equation}}
{up to the conventional scheme-dependent prefactor \cite{Gross:1973id,Politzer:1973fx}. The light trace-channel mass, heavier glueball gap, and vacuum energy may then be written as $m=c_0\Lambda_h$, $m_{\rm gap}=c_1\Lambda_h$, and $|e_{\rm vac}|=c_eN_h^2\Lambda_h^4$, with $c_0,c_1,c_e$ fixed by the microscopic theory and, in ordinary Yang--Mills, accessible to lattice calculation \cite{Morningstar_1999,Lucini:2004my,Lucini:2010nv}. In the effective analysis below we use the equivalent MS matching variables. The MS matching gives}
{
\begin{equation}
 m=2\mu\sqrt{A},\qquad |e_{\rm vac}|=\frac{A\mu^4}{4}.
\label{eq:mGapMap}
\end{equation}}

{If $m_{\rm gap}=\zeta m$ denotes the onset of heavier scalar glueball states, with $\zeta\gtrsim1$ in a conservative single-resonance estimate, single-field control requires}
{
\begin{equation}
H_* \ll m_{\rm gap},\qquad m_{\rm eff}(\varphi_*) \ll m_{\rm gap}.
\label{eq:confCutoff}
\end{equation}}

{Here $m_{\rm eff}^2\equiv U_{,\phi\phi}(\phi_*)$ is the light adiabatic mass in the Einstein frame. During slow roll, $m_{\rm eff}^2=3\eta_*H_*^2$ up to slow-roll corrections, so the second inequality in \eqref{eq:confCutoff} is automatically weaker than the first whenever $|\eta_*|<1$ and $H_*\ll m_{\rm gap}$. The ratio $H_*/m$ is therefore a transparent lower-bound diagnostic of the separation from the full confining spectrum. For an ordinary Yang--Mills-like spectrum $\zeta=O(1)$ and the conservative requirement is $H_*/m\ll1$. In a near-conformal hidden sector, however, the dilatonic scalar can be parametrically lighter than the heavier trace-channel resonances; then $\zeta\gg1$ and the relevant condition is $H_*/(\zeta m)\ll1$ rather than $H_*/m\ll1$ itself. This distinction is important because the MS scalar mass in Eq.~\eqref{eq:mGapMap} is the light mode retained in the EFT, whereas the cutoff is set by the first omitted state. Once $(A,\mu)$ and a microscopic realization of $\zeta$ are specified, Eq.~\eqref{eq:confCutoff} becomes a genuine self-consistency constraint rather than a free assumption.}

\begin{table}[t]
\centering
\begin{tabular}{cccccc}
\toprule
$A/\lambda$ & $\xi$ & $n_s$ & $r$ & $H_*/\Lambda_{\rm bg}$ & $H_*/m$\\
\midrule
$10^{-4}$ & $4.72\times10^3$ & $0.9651$ & $3.52\times10^{-3}$ & $4.15\times10^{-4}$ & $0.736$\\
$10^{-3}$ & $4.59\times10^3$ & $0.9663$ & $3.78\times10^{-3}$ & $4.24\times10^{-4}$ & $0.241$\\
$5\times10^{-3}$ & $4.10\times10^3$ & $0.9706$ & $5.00\times10^{-3}$ & $4.61\times10^{-4}$ & $0.124$\\
$10^{-2}$ & $3.67\times10^3$ & $0.9747$ & $6.68\times10^{-3}$ & $5.04\times10^{-4}$ & $0.101$\\
\bottomrule
\end{tabular}
\caption{{$A_s$-matched benchmarks at $N=55$ for $\lambda=10^{-2}$ and $\mu=10^{16}\,{\rm GeV}$. The gravitational cutoff is safely above $H_*$ throughout the displayed range. The intrinsic confining-sector ratio $H_*/m$ is more restrictive and should be read as a diagnostic of the boundary of validity: for $\zeta\simeq1$, entries with $H_*/m=O(1)$ sit outside the conservative single-field domain, while the hierarchy improves for larger $A$, larger $\mu$, or heavier-resonance separation $m_{\rm gap}/m=\zeta$.}}
\label{tab:cutoffbench}
\end{table}

\begin{figure}[!htbp]
\centering
\includegraphics[width=0.92\linewidth]{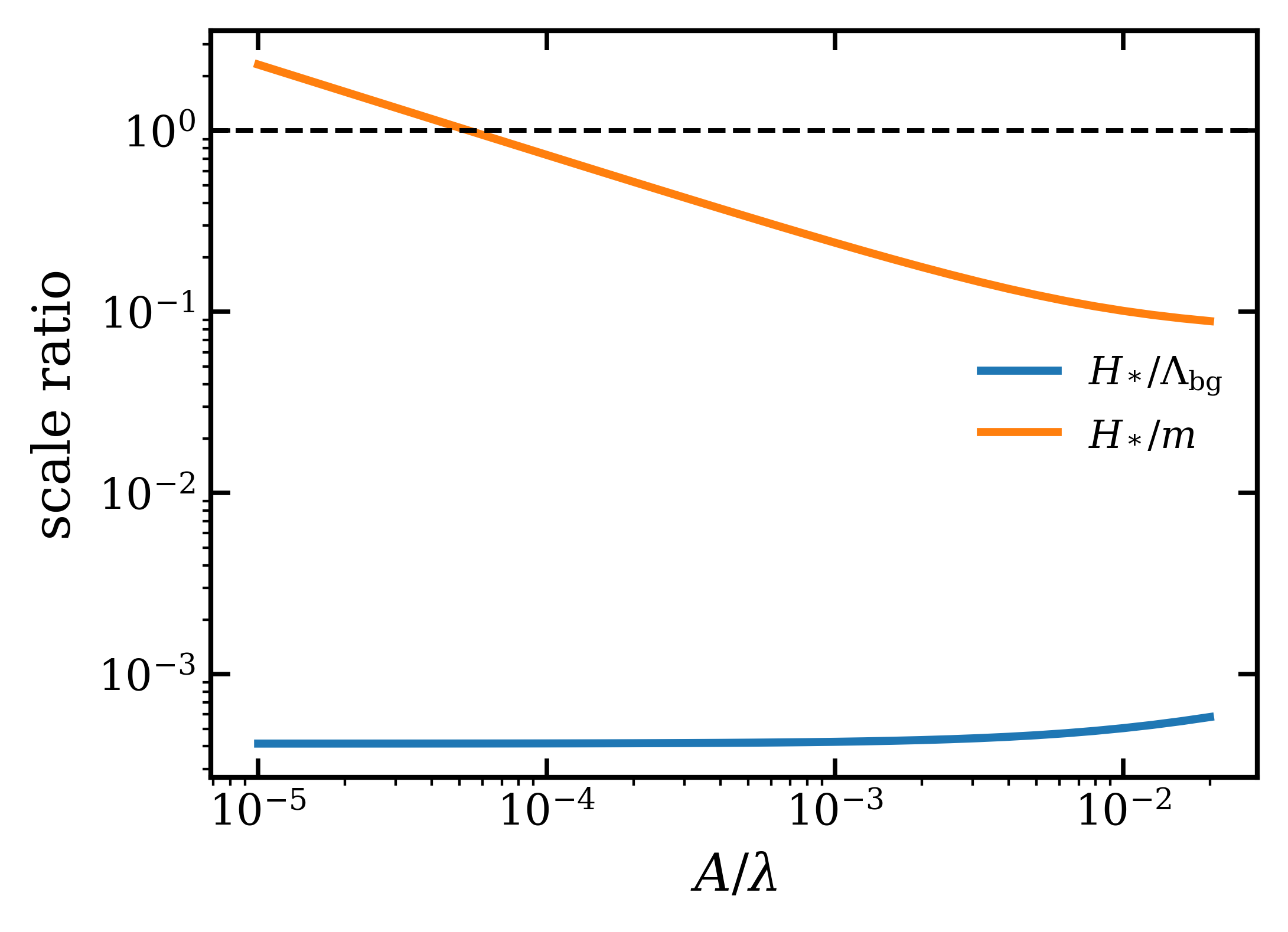}
\caption{{EFT hierarchy on the $A_s$-matched locus for $\lambda=10^{-2}$, $\mu=10^{16}\,{\rm GeV}$, and $N=55$. The nonminimal background scale is safely above $H_*$. The intrinsic confining-sector ratio $H_*/m$ is the limiting diagnostic and must be imposed together with the CMB constraints; the line $H_*/m=1$ marks the edge of decoupling, while a conservative single-field regime requires $H_*/m\ll1$ or $H_*/m_{\rm gap}\ll1$.}}
\label{fig:cutoffConsistency}
\end{figure}

{The benchmark $\mu=10^{16}\,{\rm GeV}$ is deliberately conservative because the MS relation $m=2\mu\sqrt A$ makes the hierarchy sensitive to $\mu$. For example, at $\lambda=10^{-2}$, $A/\lambda=5\times10^{-3}$ and $\mu=3\times10^{17}\,{\rm GeV}$, the same CMB normalization gives $H_*/m\simeq4.2\times10^{-3}$ and $H_*/\Lambda_{\rm bg}\simeq4.6\times10^{-4}$. This demonstrates controlled parameter space obtained by imposing the confining-sector gap condition explicitly rather than assuming it.}

{The large-$N_h$ estimate \eqref{eq:largeNMSscaling} provides a useful way to read this requirement. Taking the order-one coefficients $c_e$ and $c_0$ to be fixed by the hidden gauge dynamics, increasing $N_h$ simultaneously raises $\mu/m$ and lowers $A$. Near-conformal dynamics provides a complementary route by increasing $\zeta=m_{\rm gap}/m$. The paper therefore does not rely on QCD lattice masses as dimensional input; QCD and large-$N_h$ lattice studies are used to motivate the existence and ordering of glueball resonances, while the inflationary scale is set by a separate hidden-sector matching. A fully specified hidden gauge theory would determine $(c_e,c_0,\zeta)$ nonperturbatively, but the consistency conditions used here are the necessary matching conditions that such a theory must satisfy.}

\paragraph{Reheating.}
As a hidden-sector composite scalar, reheating proceeds through portal operators (e.g.\ couplings of the trace channel to Standard Model fields, or Higgs/curvature portals). These can be chosen weak enough not to affect slow roll while still enabling efficient reheating; consequently, reheating is model dependent and largely decoupled from the anomaly-matching origin of $V_{\rm MS}$ and the nonminimal flattening mechanism, and we leave a detailed treatment for future work.
\section{Conclusions}
\label{sec:conclusion}
% ==========================================================

{We have presented a single-field inflationary model in which the inflaton is a dilatonic scalar tied to the trace anomaly of a hidden confining gauge sector. Starting from the Migdal--Shifman anomaly-matching Lagrangian \eqref{eq:LMS}--\eqref{eq:VMSX}, we derived the canonical four-dimensional potential and the exact map $A=m^4/(64|e_{\rm vac}|)$, $\mu=4\sqrt{|e_{\rm vac}|}/m$. This establishes the nonperturbative origin of the logarithmic contribution and distinguishes it from a generic Coleman--Weinberg insertion.}

{The gravitationally completed model contains two additional Wilson coefficients, $\lambda$ and $\xi$. We have made their status explicit and have shown why they are physically needed: the pure MS limit is an important baseline and identifies the regime selected by CMB compatibility and EFT control, whereas the nonminimal coupling produces the plateau and the MS term supplies a calculable logarithmic deformation. The shift $\mu\to\mu e^{\delta}$ is absorbed by $\lambda\to\lambda-4A\delta$ at fixed $A$, but the local slope coefficient $A$ remains physical once the MS matching is specified. Neglect of the running of $\lambda$ and $\xi$ is justified only under the finite-window conditions \eqref{eq:RGcontrol}; in the plateau regime the CMB interval corresponds to the small logarithmic range \eqref{eq:CMBlogwindow}, making the fixed-coupling treatment a controlled pivot-scale approximation whenever the beta functions are moderate.}

{Using exact Einstein-frame slow-roll expressions, we found the expected large-$\xi$ attractor behavior, $n_s\simeq1-2/N$ and $r\simeq12/N^2$, together with controlled departures governed by $A/\lambda$ and $\ln(\varphi_*/\mu)$. We also imposed both EFT cutoffs: the background-dependent nonminimal scale $\Lambda_{\rm bg}$ and the intrinsic confining-sector scale $m_{\rm gap}$. Viable benchmarks satisfy $H_*/\Lambda_{\rm bg}\ll1$ and can satisfy $H_*/m_{\rm gap}\ll1$ with explicit choices of $(A,\mu)$ consistent with the MS map; large-$N_h$ scaling or near-conformal resonance separation supplies concrete hidden-sector mechanisms for the required hierarchy. Points that fail the latter hierarchy are retained in the diagnostic plots to display the edge of the single-field MS description, with the controlled domain selected by $H_*/m_{\rm gap}\ll1$. The resulting scenario is therefore a nonminimal plateau inflation model whose leading logarithmic imprint is anomaly anchored, microscopically interpretable, and quantitatively constrained by CMB observables and EFT consistency.}

\FloatBarrier
\begin{acknowledgments}
I.K. acknowledges support from Zhejiang Normal University through a postdoctoral fellowship under Grant No.~YS304224924. TL is supported in part by the National Key Research and Development Program of China Grant No. 2020YFC2201504, by the Projects No. 11875062, No. 11947302, No. 12047503, and No. 12275333 supported by the National Natural Science Foundation of China, by the Key Research Program of the Chinese Academy of Sciences, Grant No. XDPB15, by the Scientific Instrument Developing Project of the Chinese Academy of Sciences, Grant No. YJKYYQ20190049, by the International Partnership Program of Chinese Academy of Sciences for Grand Challenges, Grant No. 112311KYSB20210012, and by the Henan Province Outstanding Foreign Scientist Studio Project, No.GZS2025008.

\end{acknowledgments}

\bibliography{refs}

\end{document}